\documentclass[%
 reprint,
superscriptaddress,
 amsmath,amssymb,
 aps,
 prd,
 floatfix
]{revtex4-2}

\usepackage{graphicx}
\usepackage{dcolumn}
\usepackage{bm}
\usepackage{slashed}
\usepackage{url}
\usepackage{hyperref}
\usepackage{xcolor}
\usepackage{comment}
\usepackage{enumitem}
\usepackage{blkarray}
\usepackage{physics}
\usepackage{mathtools} 
\AtBeginDocument{%
  \setlength\abovedisplayskip{9pt plus 2pt minus 2pt}%
  \setlength\belowdisplayskip{9pt plus 2pt minus 2pt}%
  \setlength\abovedisplayshortskip{0pt plus 2pt}%
  \setlength\belowdisplayshortskip{7pt plus 2pt minus 2pt}%
}

\newcommand{\DeltaLP}{\ensuremath{\Delta}}
\newcommand{\deltaLP}{\ensuremath{\delta}}
\newcommand{\deltamP}{\ensuremath{\hat{m}_{\mu}}}

\begin{document}

\preprint{APS/123-QED}

\title{Scalar and Tensor Form Factors for
$\Lambda \rightarrow p\ell \bar{\nu}_\ell$ from Lattice QCD}

\author{Constantia Alexandrou}
\affiliation{Computation-based Science and Technology Research Center,
The Cyprus Institute, 20 Kavafi Str., Nicosia 2121, Cyprus}
\affiliation{Department of Physics, University of Cyprus, P.O. Box 20537, 1678 Nicosia, Cyprus}

\author{Simone Bacchio}
\affiliation{Computation-based Science and Technology Research Center,
The Cyprus Institute, 20 Kavafi Str., Nicosia 2121, Cyprus}
\author{Andreas Konstantinou}
\affiliation{Computation-based Science and Technology Research Center,
The Cyprus Institute, 20 Kavafi Str., Nicosia 2121, Cyprus}
\affiliation{Department of Physics, University of Cyprus, P.O. Box 20537, 1678 Nicosia, Cyprus}
\author{Eleni Vakana}
\affiliation{Department of Physics, University of Cyprus, P.O. Box 20537, 1678 Nicosia, Cyprus}

\begin{abstract}
We present a determination of the scalar and tensor $\Lambda\to p$ transition form factors using lattice QCD. These form factors are  relevant for semileptonic hyperon decays  in the presence of extensions of  the Standard Model that include scalar and tensor interactions. The calculation is carried out using a gauge ensemble of twisted mass fermions at the physical pion mass, following the same strategy as our recent study on vector and axial form factors for the same transition~\cite{Bacchio:2025auj}. We provide the complete set of form factors as functions of $q^2$ employing a model-independent parametrization.
We examine their impact on searches for non-standard charged-current interactions via the muon-to-electron decay-rate ratio $R^{\mu e}=\Gamma(\Lambda\to p\mu\bar\nu_\mu)/\Gamma(\Lambda\to pe\bar\nu_e)$, where scalar and tensor contributions enter linearly and are helicity-enhanced relative to the electron channel. We compare this first-principles prediction for the decay-rate ratio with recent experimental measurements, thereby enabling improved constraints on non-standard charged-current interactions.
\end{abstract}


\maketitle


\section{Introduction}
\label{sec:intro}

Semileptonic decays of hyperons provide a sensitive and complementary laboratory for testing the structure of charged-current interactions. In the Standard Model (SM), $\Delta S=1$ transitions such as $\Lambda\to p\,\ell\,\bar\nu_\ell$ are related to the left-handed $V\!-\!A$ quark current and are described, at low energies, by a set of six hadronic form factors that encode the underlying non-perturbative QCD dynamics. These decays have long played an important role in the determinations of the Cabibbo--Kobayashi--Maskawa (CKM) \cite{Cabibbo:2003cu,Kobayashi:1973fv} matrix element $V_{us}$ and in tests of flavor symmetries, where the normalization of the vector form factor at $q^2=0$ is protected by the Ademollo--Gatto theorem against first-order $SU(3)$-breaking effects \cite{Ademollo:1964sr}. In this context, quantitative control over the hadronic matrix elements is essential for precision phenomenology.

Beyond their role in SM physics, semileptonic hyperon decays  also provide a window on possible new-physics contributions to charged-current interactions \cite{Cirigliano:2013xha,Chang:2014iba,Cirigliano:2018dyk}. In a model-independent effective-field-theory description at energies well below the electroweak scale, additional scalar and tensor operators can accompany the SM left-handed $(V{-}A)$ structure in $s\to u\,\ell\bar\nu_\ell$ transitions~\cite{Cirigliano:2009wk,Chang:2014iba,Gonzalez-Alonso:2016etj,Becirevic:2026tle}. Such interactions are strongly constrained by precision measurements in nuclear and neutron $\beta$ decay, pion and kaon decays, and by high-$p_T$ searches at the LHC \cite{Bhattacharya:2011qm,Gonzalez-Alonso:2016etj,Gonzalez-Alonso:2018omy}, yet remain phenomenologically interesting because they can modify kinematic distributions and enter certain observables linearly through interference with the SM amplitude.

A particularly clean probe is the muon-to-electron decay-rate ratio $R^{\mu e}=\Gamma(\Lambda\to p\mu\bar\nu_\mu)/\Gamma(\Lambda\to pe\bar\nu_e)$. This ratio is independent of $V_{us}$ and the lifetime of $\Lambda$ and, at next-to-leading order in the SM, is free of hadronic form-factor uncertainties \cite{Chang:2014iba}. As emphasized in Ref.~\cite{Chang:2014iba}, scalar and tensor interactions generate parametrically enhanced contributions to $R^{\mu e}$ through the muon-mass dependence of the interference terms, making this ratio an efficient discriminator of non-standard charged currents.

Realizing the full potential of semileptonic hyperon decays as precision probes requires reliable non-perturbative determinations of the relevant hadronic input. Lattice QCD provides a systematically improvable framework to compute these form factors from first principles. Over the past decade, substantial progress  has been made in controlling the dominant lattice systematics, in particular, in controlling excited-state contamination and in computing  matrix elements using  gauge ensembles simulated at the physical quark masses avoiding uncontrolled systematics from chiral extrapolations. In addition,  using multiple ensembles to perform the continuum limit  has enabled precision determinations of nucleon charges, axial form factors, and electromagnetic form factors \cite{Walker-Loud:2019cif,Bali:2023sdi,Djukanovic:2024krw,Alexandrou:2024ozj,FlavourLatticeAveragingGroupFLAG:2024oxs,Jang:2023zts,Djukanovic:2022wru,Alexandrou:2023qbg,Djukanovic:2023beb,Tsuji:2023llh,Alexandrou:2025vto}.
These developments motivate analogous high-precision calculations for semileptonic baryon decays, where form factors provide the essential non-perturbative input. 

Lattice QCD studies of semileptonic decays of heavy baryons have reached a mature stage, including, for example, calculations of $\Lambda_b\to N$~\cite{Detmold:2015aaa}, $\Lambda_c\to\Lambda$ \cite{Meinel:2016dqj}, $\Lambda_c\to N$ \cite{Meinel:2017ggx}, $\Lambda_c\to \Lambda^{(1520)}$ \cite{Meinel:2021mdj}, $\Lambda_b\to \Lambda_c^{(2595,2625)}$ \cite{Meinel:2021rbm}, and $\Xi_c\to\Xi$ \cite{Farrell:2025gis}. In contrast, lattice results for hyperon semileptonic transitions remain comparatively limited, with existing calculations focusing on selected vector form factors in $\Sigma\to N$ and $\Xi\to\Sigma$ channels \cite{Shanahan:2015dka,Sasaki:2017jue} or on earlier studies at heavier than physical pion masses \cite{Sasaki:2008ha}. In our recent lattice-QCD study~\cite{Bacchio:2025auj}, the vector and axial-vector $\Lambda\to p$ form factors were determined using the same  $N_f=2+1+1$  gauge ensemble with pion mass tuned to its physical value as the one analyzed in this work. It enabled a fully non-perturbative prediction for the SM decay rate and for the ratio $R^{\mu e}$.  In this work, we compute the scalar and tensor $\Lambda \to p$ transition form factors using the same  analysis strategy providing their full $q^2$ dependence through a model-independent parametrization. These results provide the missing non-perturbative input needed to sharpen semileptonic hyperon decays  constraints on scalar and tensor interactions and to update the $\Lambda\to p$ contribution in the global semileptonic hyperon decays analysis of Ref.~\cite{Chang:2014iba}. 

The rest of the paper is organized as follows: In Sec.~\ref{sec:Thry}, we summarize the low-energy effective description of $s\to u\,\ell\bar\nu_\ell$ transitions in beyond the SM theories and define the scalar and tensor form factors and their connection to $R^{\mu e}$. Sec.~\ref{sec:lattice} describes the lattice setup and the extraction of the scalar and tensor matrix elements, including the control of excited-state effects and the parametrization of the $q^2$ dependence. The lattice QCD results for the form factors obtained and their phenomenological implications are presented in Sec.~\ref{sec:results}. We conclude in Sec.~\ref{sec:concl}.


\section{Theoretical Formalism}
\label{sec:Thry}

At energies well below the electroweak scale, semileptonic $s {\to} u\,\ell\bar\nu_\ell$ transitions are described within an effective field theory framework. Allowing for non-standard charged-current interactions, the leading-order low-energy effective Lagrangian including scalar and tensor operators reads~\cite{Chang:2014iba}

\begin{align}
\mathcal{L}_{\rm eff} = -\frac{G_F V_{us}}{\sqrt{2}}
\sum_{\ell=e,\mu}\Big[
\bar\ell\gamma_\mu(1{-}\gamma_5)\nu_\ell\;
\bar u\gamma^\mu(1{-}\gamma_5)s+
\nonumber\\
\epsilon_S\,
\bar\ell(1{-}\gamma_5)\nu_\ell\;
\bar u s
- \epsilon_T\,
\bar\ell\sigma_{\mu\nu}(1-\gamma_5)\nu_\ell\;
\bar u\sigma^{\mu\nu}(1{-}\gamma_5)s
\Big] + \mathrm{h.c.}\,,
\label{eq:Leff}
\end{align}
where $\sigma^{\mu\nu}=\tfrac{i}{2}[\gamma^\mu,\gamma^\nu]$, $\epsilon_{S,T}$ are dimensionless coefficients parameterizing possible physics beyond the SM, and $V_{us}$ is the CKM matrix element. The SM limit is recovered for $\epsilon_S=\epsilon_T=0$. More general extensions of the SM also generate additional operators, such as right-handed currents or pseudoscalar interactions~\cite{Chang:2014iba}. These are not considered here since right-handed currents do not introduce new Lorentz structures, while pseudoscalar contributions are further suppressed by kinematics and by the charged-lepton mass in interference terms with SM amplitudes. By contrast, scalar and tensor operators produce genuinely new Dirac structures and distinct interference patterns in observables such as $R^{\mu e}$.
Accordingly, the lattice-QCD form factors computed in this work fix the hadronic matrix elements arising from the quark bilinears in Eq.~\eqref{eq:Leff}, allowing controlled predictions for observables sensitive to tensor and scalar interactions.
\\
\subsection{Decay rate and the $R^{\mu e}$ observable}

The decay rate of the semileptonic process
$\Lambda \to p\,\ell\,\bar\nu_\ell$ is given by
\begin{multline}\label{eq:decay_rate}
\Gamma = \frac{1}{2m_\Lambda}
\int \frac{d^3\vec{p}_p}{(2\pi)^3 2E_p}
      \frac{d^3\vec{p}_\ell}{(2\pi)^3 2E_\ell}
      \frac{d^3\vec{p}_\nu}{(2\pi)^3 2E_\nu}
\\ \times
(2\pi)^4 \delta^{(4)}(p_\Lambda - p_p - p_\ell - p_\nu)
\,\big|\mathcal{M}\big|^2 ,
\end{multline}
where $\mathcal{M}$ denotes the invariant matrix element, $p$ is the four-momentum of each particle, and $E=\sqrt{m^2+|\vec p\,|^2}$.
After integrating over the angular variables, the decay rate depends only on the squared momentum transfer $q^2=(p_\Lambda-p_p)^2=(p_\ell+p_\nu)^2$.

To probe non-standard interactions, it is convenient to consider the
ratio of muonic to electronic decay rates,
\begin{equation}
R^{\mu e}\equiv
\frac{\Gamma(\Lambda\to p\,\mu\,\bar\nu_\mu)}
     {\Gamma(\Lambda\to p\,e\,\bar\nu_e)}.
\end{equation}
This observable is particularly clean, as it is independent of the CKM matrix element and the Fermi constant $G_F$, allowing for a genuine first-principles prediction via lattice QCD calculations. In addition, it can be shown~\cite{Chang:2014iba} that this ratio is independent of hadronic form factors up to percent-level effects, namely
\begin{align}\label{eq.Rat}
R^{\mu e}
&=
\sqrt{1-\deltamP^2}
\left(1-\frac{9}{2}\deltamP^2
      -4\deltamP^4\right)
\\
&\quad
+\frac{15}{2}\deltamP^4
\operatorname{artanh}
\left(\sqrt{1-\deltamP^2}\right)
+\mathcal{O}\!\left(\deltaLP^2,\epsilon_S,\epsilon_T\right)\,,\nonumber
\end{align}
with
\begin{equation}\label{eq:params}
    \deltamP \equiv \frac{m_\mu}{\DeltaLP}\,,\quad \deltaLP \equiv \frac{\DeltaLP}{m_\Lambda}\,,\ \text{and}\ \DeltaLP = m_\Lambda - m_p\,,
\end{equation}
and where $\mathcal{O}(\epsilon_S,\epsilon_T)$ denotes corrections arising from
non-standard scalar and tensor interactions that will be introduced in the following.

\subsection{Scalar and tensor form factors}

The vector, axial-vector, scalar, and tensor matrix elements relevant for the $\Lambda\to p$ transition are parameterized in terms of Lorentz-invariant form factors in Minkowski space-time $g_{\mu\nu}=\texttt{diag(+,-,-,-)}$ as 
\begin{widetext}
\begin{align}
\langle p(p_p,s_p) | \bar u\,\gamma_\mu s | \Lambda(p_\Lambda,s_\Lambda)\rangle  &= 
\bar{u}_p \Big[ \gamma_\mu F_1(q^2)  -i \sigma_{\mu\nu} \frac{q^\nu}{m_{\Lambda}}F_2(q^2)    + \frac{q_\mu}{m_{\Lambda}} F_3(q^2) \Big] u_{\Lambda}\,,\\
\langle p(p_p,s_p) | \bar u\,\gamma_\mu\gamma_5 s | \Lambda(p_\Lambda,s_\Lambda)\rangle  &= 
\bar{u}_p \Big[ \gamma_\mu G_1(q^2)  -i \sigma_{\mu\nu} \frac{q^\nu}{m_{\Lambda}}G_2(q^2)    + \frac{q_\mu}{m_{\Lambda}} G_3(q^2) \Big]\gamma_5 u_{\Lambda}\,,\\
\langle p(p_p,s_p) | \bar u\, s | \Lambda(p_\Lambda,s_\Lambda)\rangle&=
F_S(q^2)\,\bar u_p\,u_\Lambda\,,\label{eq:scalar_decomp}\\
\langle p(p_p,s_p)|\bar u\,i\sigma^{\mu\nu} s|\Lambda(p_\Lambda,s_\Lambda)\rangle
 &=\bar u_p\Big[
i\sigma^{\mu\nu}T_1(q^2)\,+\Sigma^{\mu\nu}_{\sigma\rho}\Big(
\gamma^\sigma \frac{ p_\Lambda^\rho}{m_\Lambda}T_2(q^2)
+\gamma^\sigma\frac{p_p^{\rho}}{m_p} T_3(q^2)
+\frac{p_\Lambda^\sigma p_p^{\rho}}{m_\Lambda m_p}T_4(q^2)
\Big)\Big]u_\Lambda
\label{eq:tensor_decomp}\,,
\end{align}
\end{widetext}
where $\Sigma^{\mu\nu}_{\sigma\rho}=g^\mu_\sigma g^\nu_\rho-g^\nu_\sigma g^\mu_\rho$, the Bjorken Drell \cite{Bjorken:1965zz} conventions for the Dirac matrices is used, and
$u_{\Lambda,p}$ are relativistically normalized Dirac spinors,
\begin{equation}
u_B=\frac{\slashed{p}_B+m_B}{\sqrt{2E_B(E_B+m_B)}}\,u^s,
\qquad B\in\{p,\Lambda\}.
\end{equation}
The elements of $\langle p(p_p,s_p)|\bar u\,\sigma^{\mu\nu}\gamma_5 s|\Lambda(p_\Lambda,s_\Lambda)\rangle$ are obtained by using $\sigma^{\mu\nu}\gamma_5=-i\epsilon^{\mu \nu \lambda \rho}\sigma_{\lambda \rho}$.

For phenomenological applications to semileptonic hyperon decays, it is common to retain only the leading tensor structure, absorbing subleading terms suppressed by $q/m_\Lambda$ into a single effective tensor form factor~\cite{Chang:2014iba}, defining
\begin{equation}
\langle p|\bar u\, i\sigma^{\mu\nu} s|\Lambda\rangle
=
F_T(q^2)\,
\bar u_p\,i\sigma^{\mu\nu}u_\Lambda
+\mathcal{O}(q/m_\Lambda)\,.
\label{eq:fT_def}
\end{equation}
Here, instead, we employ the full decomposition in Eq.~\eqref{eq:tensor_decomp} and therefore identify  $F_T(q^2)\simeq T_1(q^2)$ in our analysis for comparison purposes. The primary focus of this work is the determination of the scalar $F_S(q^2)$ and tensor $T_{1,2,3,4}(q^2)$ form factors. For more details on the vector $F_{1,2,3}(q^2)$ and axial-vector $G_{1,2,3}(q^2)$ form factors and their consistent determination, we refer to our recent work in Ref.~\cite{Bacchio:2025auj}.

\subsection{Sensitivity to scalar and tensor interactions}

The differential decay rate in presence of non-standard interactions, is given by \cite{Liu:2023zvh}
\begin{equation}
\frac{d\Gamma^\ell}{dq^2}=\frac{d\Gamma^\ell_{LL}}{dq^2}+\epsilon_S\frac{d\Gamma^\ell_{LS}}{dq^2}+\epsilon_T\frac{d\Gamma^\ell_{LT}}{dq^2}+\mathcal{O}(\epsilon_S^2,\epsilon_T^2)\,,
    \label{eqn:dif}
\end{equation}
where the various terms arise from expanding the squared amplitude $|\mathcal{M}|^2$ in Eq.~\eqref{eq:decay_rate}. Here $\Gamma^\ell_{LL}$ denotes the purely SM left-handed ($L$) squared-amplitude contribution, while $\Gamma^\ell_{LS}$ and $\Gamma^\ell_{LT}$ are the leading-order non-standard scalar ($S$) and tensor ($T$) terms interfering with the left-handed interaction. Higher-order contributions in $\epsilon_{S}^2$ and $\epsilon_{T}^2$ correspond to the purely scalar and tensor squared-amplitude terms, $\Gamma^\ell_{SS}$ and $\Gamma^\ell_{TT}$. These are neglected because they are higher order in $\epsilon$. Further details on this point can be found in the Appendix of Ref.~\cite{Liu:2023zvh}.
The explicit expressions for the decay rates, after integrating over the angular degrees of freedom, are defined in terms of helicity vector $F_{0,+,\perp}(q^2)$ and axial $G_{0,+,\perp}(q^2)$ form factors as \cite{Liu:2023zvh},
\begin{widetext}
\begin{align}
\frac{d\Gamma^\ell_{LL}}{dq^2}&=\frac{G_F^2 |V_{us}|^2 \sqrt{s_+ s_-}}{192 \pi^3 m_\Lambda^3}\left(1-\frac{m_\ell^2}{q^2}\right)^{\!2}s_+ (m_\Lambda-m_p)^2\Bigg\{\frac{3m_\ell^2}{2q^2} \bigg[\frac{s_- (m_\Lambda+m_p)^2}{s_+ (m_\Lambda-m_p)^2} G_0^2(q^2)+ F_0^2(q^2)\bigg]
\nonumber\\&+\frac{m_\ell^2+2q^2}{2q^2}\bigg[G_+^2(q^2)+\frac{s_- (m_\Lambda+m_p)^2}{s_+ (m_\Lambda-m_p)^2} F_+^2(q^2)\bigg]
+\frac{m_\ell^2+2q^2}{(m_\Lambda-m_p)^2}\bigg[G_\perp^2(q^2)+\frac{s_-}{s_+} F_\perp^2(q^2)\bigg]\Bigg\},\\[6pt]
\frac{d\Gamma^\ell_{LS}}{dq^2}&=\frac{G_F^2 |V_{us}|^2 \sqrt{s_+ s_-}}{64\pi^3 m_\Lambda^3}\left(1-\frac{m_\ell^2}{q^2}\right)^{\!2}m_\ell s_+ (m_\Lambda-m_p)F_S(q^2)F_0(q^2),\label{eq:decay_rate_LS}
\\[6pt]
\frac{d\Gamma^\ell_{LT}}{dq^2}&=\frac{G_F^2 |V_{us}|^2 \sqrt{s_+ s_-}}{16\pi^3 m_\Lambda^3}\left(1-\frac{m_\ell^2}{q^2}\right)^{\!2}m_\ell s_+
\Bigg\{(m_\Lambda-m_p)\,T_1(q^2) \bigg[2\frac{s_-(m_\Lambda+m_p)}{s_+(m_\Lambda-m_p)}F_\perp(q^2)-2G_\perp(q^2)-G_+(q^2)\bigg]
\nonumber\\
&-\frac{s_-}{m_\Lambda}
T_2(q^2)
\bigg[
\frac{m_\Lambda^2-m_p^2+q^2}{s_+}F_\perp(q^2)
-G_\perp(q^2)
\bigg]-\frac{s_-}{m_p}T_3(q^2)\bigg[\frac{m_\Lambda^2-m_p^2-q^2}{s_+}F_\perp(q^2)-G_\perp(q^2)\bigg]\Bigg\}.
\end{align}
\end{widetext}
with $s_{\pm}=(m_\Lambda\pm m_p)^2-q^2$. The mapping of the helicity form factors to the Weinberg form factors $F_{1,2,3}(q^2)$ and $G_{1,2,3}(q^2)$ can be found in Eqs.~(21--26) of Ref.~\cite{Bacchio:2025auj}. 

After integrating Eq.~\eqref{eqn:dif} over $q^2$,
\begin{equation}\label{eq:decay_int}
    \Gamma^\ell_i = \int_{m_\ell^2}^{q_{\max}^2}\frac{d\Gamma^\ell_i}{dq^2}\, dq^2 \qq{with} q_{\max}^2=(m_\Lambda-m_p)^2
\end{equation}
the decay rate of $\Lambda \rightarrow p\ell \bar{\nu}_\ell$ is given by 
\begin{align}\label{eq:decay_rate_decomp}
&\Gamma^\ell=\Gamma^\ell_{\rm SM}+\Gamma^\ell_{\rm NS}\qq{with} \Gamma^\ell_{\rm SM} = \Gamma^\ell_{LL}\nonumber\\ &\qq{and}\Gamma^\ell_{\rm NS} = \epsilon_S\Gamma^\ell_{LS}+\epsilon_T \Gamma^\ell_{LT}+\mathcal{O}(\epsilon^2),
\end{align}
where the SM is only LL, i.e. $\epsilon=0$, while the non-standard~(NS)  part contains scalar and tensor interactions.
Following the analysis of Ref.~\cite{Chang:2014iba}, the ratio
$R^{\mu e}/R^{\mu e}_{\rm SM}$ can be expressed as
\begin{align}
\frac{R^{\mu e}}{R^{\mu e}_{\rm SM}}=&\frac{\Gamma^e_{\rm SM}}{\Gamma^\mu_{\rm SM}}\frac{(\Gamma^\mu_{\rm SM}+\Gamma^\mu_{\rm NS})}{(\Gamma^e_{\rm SM}+\Gamma^e_{\rm NS})}
 \simeq
1+r_S\,\epsilon_S+r_T\,\epsilon_T+\mathcal{O}(\epsilon^2)\,,\nonumber\\
\qq{with}
&r_S \equiv \frac{1}{{R^{\mu e}_{\rm SM}}}\frac{\partial R^{\mu e}}{\partial\epsilon_S}\Bigg|_{\epsilon=0} \!\!=\frac{\Gamma^\mu_{LS}}{\Gamma^\mu_{LL}}-\frac{\Gamma^e_{LS}}{\Gamma^e_{LL}},\nonumber\\
\qq{and}&r_T \equiv\frac{1}{{R^{\mu e}_{\rm SM}}}\frac{\partial R^{\mu e}}{\partial\epsilon_T}\Bigg|_{\epsilon=0}\!\!=\frac{\Gamma^\mu_{LT}}{\Gamma^\mu_{LL}}-\frac{\Gamma^e_{LT}}{\Gamma^e_{LL}}
\label{eq:Rlin}
\end{align}
which quantify the sensitivity to scalar and tensor interactions, respectively. At leading order, the sensitivity coefficients are

\begin{align}
\label{Eq:sens}
    r_SR^{\mu e}_{\rm SM} &= \frac{\Gamma^\mu_{LS}}{\Gamma^e_{LL}} + \mathcal{O}(m_e) = \frac{f_1f_S
}
{f_1^2+3 g_1^2}
\,
\Pi(\deltamP)+\mathcal{O}\!\left(\deltaLP^2\right),\\
    r_TR^{\mu e}_{\rm SM} &=  \frac{\Gamma^\mu_{LT}}{\Gamma^e_{LL}} + \mathcal{O}(m_e) = \frac{12
g_1f_T
}
{f_1^2+3g_1^2}
\,
\Pi(\deltamP)+\mathcal{O}\!\left(\deltaLP^2\right),\nonumber
\end{align}
where $g_i\equiv G_i(q^2=0)$ and $f_i\equiv F_i(q^2=0)$, and the phase-space function
$\Pi(\deltamP)$ is given by
\begin{multline}
\Pi(\deltamP)
=
\frac{5}{2}\deltamP
\Bigg[
\left(2+13\deltamP^2\right)
\sqrt{1-\deltamP^2}
\\
-\deltamP^2\left(4
+\deltamP^2\right)
\operatorname{artanh}
\left(\sqrt{1-\deltamP^2}\right)
\Bigg].
\label{eq:Pi}
\end{multline}
Eq.~\eqref{Eq:sens} differs by a factor of ($1{-}\tfrac{3}{2}\deltaLP$) from Eq.~(10) in Ref.~\cite{Chang:2014iba}. This is due to the fact that in this work we expand both the numerator and the denominator to order $\mathcal{O}\!\left(\deltaLP^2 \right)$, while in Ref.~\cite{Chang:2014iba} only the denominator was expanded to $\mathcal{O}\!\left(\deltaLP^2 \right)$ and the numerator at $\mathcal{O}\!\left(\deltaLP \right)$. More details on the derivation of the expansion can be found in Appendix \ref{app:sensitivity}.

\section{Lattice QCD methodology}
\label{sec:lattice}

The gauge ensemble analyzed in this work was generated by the Extended Twisted Mass Collaboration (ETMC) using the
twisted mass fermion formulation, which at maximal twist provides
automatic $\mathcal{O}(a)$ improvement of physical observables
\cite{Frezzotti:2000nk,Frezzotti:2003ni}. To reduce the isospin-breaking
effects, the fermion action
includes a clover term \cite{Sheikholeslami:1985ij}. The bare light-quark
mass parameter $\mu_l$ is tuned such that the isosymmetric pion mass is equal to its physical value, $m_\pi = 135$ MeV
\cite{Alexandrou:2018egz,Finkenrath:2022eon}. The strange and charm mass
parameters, $\mu_s$ and $\mu_c$, are fixed using the kaon mass together
with a reference ratio involving the $D$-meson mass and decay constant,
and by imposing a phenomenologically motivated relation between strange
and charm masses, following the tuning strategy of
Refs.~\cite{Alexandrou:2018egz,Finkenrath:2022eon}. The 
parameters of the gauge ensemble are collected in Table~\ref{tbl:Ensembles}.

We work in the SU(2)-isosymmetric limit, in which up and down quarks are
degenerate and the proton and neutron are, therefore, treated as a single
nucleon state, $N$. Leading isospin-breaking and electromagnetic
corrections are expected to contribute at the (sub-)percent level for the
observables considered here and are neglected, consistent with the
standard treatment in current  lattice calculations of
baryon matrix elements of similar precision as the one targeted here.
As the goal of this work is a first  determination of the
$\Lambda\to p$ scalar and tensor form factors with controlled statistical
and excited-state systematics, we perform the analysis on a single ensemble with physical pion mass, with $a\simeq 0.08$~fm and $L\simeq 5$~fm. A
dedicated quantification of discretization and finite-volume effects
requires additional lattice spacings and volumes and is left to future
work. Existing nucleon structure studies on the same ensemble suggest that such effects are small at the current level of precision, aided by the $\mathcal{O}(a)$ improvement of the action \cite{Alexandrou:2024ozj,Alexandrou:2023qbg,Alexandrou:2025vto}.
The lattice spacing and mass-tuning to the isosymmetric QCD point, employed in this analysis, is taken from
Ref.~\cite{ExtendedTwistedMass:2022jpw,ExtendedTwistedMass:2024nyi}. Although determinations based on
different observables can yield slightly different results,  at finite lattice spacing due
to cutoff effects, such differences  disappear in the
continuum limit when the scale setting and tuning are performed
consistently \cite{ExtendedTwistedMass:2021gbo}.

\begin{table}[h!]
	\centering
	{\small
		\renewcommand{\arraystretch}{1.2}
		\renewcommand{\tabcolsep}{1.5pt}
    \begin{tabular}{c|c|c|c|c|c}
    \hline \hline
        Ensemble     & Acronym & $V/a^4$            & $\beta$ & $a$~[fm]    & $m_\pi$~[MeV] \\ \hline 
        cB211.072.64 & B64   & $64^3 \times 128$     & 1.778   & 0.07957(13) & 135      \\ 
         \hline \hline
        \end{tabular}}
    \caption{Parameters of the $N_f=2+1+1 $ ensemble analyzed in this work. In the first column, we give the name of
             the ensemble, in the second the abbreviated name, in the third the lattice volume, in the fourth $\beta = 6/g^2$ with $g$ the bare coupling constant, in the fifth
             the lattice spacing and in the last the pion mass. The lattice spacing and the pion mass is determined in Ref.~\cite{ExtendedTwistedMass:2022jpw}.}
    \label{tbl:Ensembles}
\end{table}

\subsection{Generation of correlation functions}
To isolate the nucleon and $\Lambda$ ground states we employ standard
local interpolating fields,
\begin{align}
\chi_N(x) &= \epsilon^{abc} \left(u^a(x)^T C\gamma_5 d^b(x)\right) u^c(x)\,,\\[0.15cm]
\chi_\Lambda(x)&=\tfrac{1}{\sqrt{6}}\epsilon^{abc}\Big[
2\big(u^a(x)^T C\gamma_5 d^b(x)\big)s^c(x)+
\\\big(u^a(x)&^T C\gamma_5 s^b(x)\big)d^c(x)
-\big(d^a(x)^T C\gamma_5 s^b(x)\big)u^c(x)
\Big]\,,\nonumber
\end{align}
where $C$ is the charge-conjugation matrix and $a,b,c$ are color indices.
The baryon two-point functions are constructed as
\begin{equation}\label{eq:2pt}
C_B(\vec{p}_B,t_{\rm s})=
\sum_{\vec{x}_s} \Tr\!\left[
P_0\,
\langle \chi_B(x_s)\,\bar{\chi}_B(0)\rangle\,
e^{-i\vec{x}_s\cdot \vec{p}_B}
\right] ,
\end{equation}
with $B\in\{N,\Lambda\}$, $x_s=(\vec{x}_s,t_{\rm s})$ and the trace is taken over Dirac indices.

For the $\Lambda\to N$ transition we compute three-point functions with
a scalar or a tensor current insertion. Denoting the insertion point by
$x_{\rm ins}=(\vec{x}_{\rm ins},t_{\rm ins})$, we form
\begin{multline}\label{eq:3ptS_def}
C^{S}_{\mu}(\vec{p}_N,\vec{p}_\Lambda;t_{\rm s},t_{\rm ins})
=
\sum_{\vec{x}_{\rm ins},\vec{x}_s}
e^{i\vec{x}_{\rm ins}\cdot(\vec{p}_\Lambda-\vec{p}_N)
-i\vec{x}_s\cdot\vec{p}_\Lambda}
\\ \times
\Tr\!\left[
P_\mu\,
\langle \chi_\Lambda(x_s)\,
\bar u(x_{\rm ins})\,s(x_{\rm ins})\,
\bar{\chi}_N(0)\rangle
\right],
\end{multline}
and, analogously, tensor three-point functions, $C_{\mu \nu \lambda}^T(\vec{p}_N,\vec{p}_\Lambda;t_{\rm s},t_{\rm ins})$, with the insertion
$\bar u(x_{\rm ins})\,\sigma_{\nu\lambda}\,s(x_{\rm ins})$ and $\mu$ being the index of the projector $P_\mu$. For
unpolarized positive-parity matrix elements we take
$P_0=\tfrac12(1+\gamma_0)$, while for polarized matrix elements we use
$P_k=iP_0\gamma_5\gamma_k$.
In this work we employ a fixed sink setup with $\vec{p}_\Lambda=\vec{0}$
and vary the nucleon momentum $\vec{p}_N$, such that the momentum
transfer satisfies $\vec{q}=\vec{p}_\Lambda-\vec{p}_N=-\vec{p}_N$.

The correlation functions required to determine matrix elements for the
$\Lambda \to N$ transition are computed at the simulated light-quark mass
$a\mu_{ud,\text{sea}} = 0.00072$ and at a strange-quark mass
$a\mu_{s,\text{isoQCD}} = 0.018267(53)$, which has been refined
\cite{ExtendedTwistedMass:2024nyi} to match the adopted isoQCD
prescription as defined by the Edinburgh/FLAG consensus
\cite{FlavourLatticeAveragingGroupFLAG:2024oxs}. The residual mismatch
between $a\mu_{ud,\text{sea}}$ and the target isoQCD value
$a\mu_{ud,\text{isoQCD}} = 0.0006669(28)$ is expected to induce effects
below the percent level, based on dedicated studies in related contexts
\cite{ExtendedTwistedMass:2022jpw,ExtendedTwistedMass:2024nyi,Alexandrou:2023dlu},
and is therefore smaller than the accuracy goal of the present work.

Since lattice matrix elements must be matched to their continuum definitions, renormalization is required. For the scalar and tensor currents we employ the RI-MOM scheme~\cite{Martinelli:1994ty}, in contrast to the hadronic method used for the axial and vector currents in Ref.~\cite{Bacchio:2025auj}.
The resulting renormalization factors $Z_P$ and $Z_T$ in $\overline{\rm MS}$ at $\mu=2\,$GeV for the B64 ensemble used here are
\begin{equation}
Z_P = 0.4864(38) \qq{ and } Z_T = 0.7948(26).
\end{equation}
They renormalize  multiplicatively  the matrix elements computed with the local operators
$S(x)=\bar{u}(x)s(x)$ and
$T_{\mu\nu}(x)=\bar{u}(x)\sigma_{\mu\nu}s(x)$, respectively. Further details on their determination will be presented in an upcoming publication.

\subsection{Smearing and statistics}

At large Euclidean time separations, baryon correlation functions
exhibit an exponentially degrading signal-to-noise ratio. To improve the
ground-state overlap and to extract it as early in time as possible, we use smeared sources at
multiple source locations per configuration, translate all sources to a
common origin ($x_0=0$), and average over them. Quark fields are smeared
in the spatial directions using Gaussian smearing
\cite{Gusken:1989ad,Alexandrou:1992ti},
\begin{equation}
q_G(\vec{x}, t) = \left[ 1 + \alpha_G \sum_{j=1}^3 \left(\vec{T}_{+j} + \vec{T}_{-j} \right) \right]^{N_G} q(\vec{x}, t),
\end{equation}
where $\alpha_G$ and $N_G$ denote the smearing parameters and $j=1,2,3$
runs over spatial directions. The gauge links entering the smearing
operator are APE-smeared \cite{APE:1987ehd}. We employ different numbers
of Gaussian smearing iterations for light and strange quarks, as listed
in Table~\ref{tab:stat}.

Table~\ref{tab:stat} also summarizes the statistics used in the two- and
three-point functions. For the three-point functions, the number of
source positions is increased with the source--sink time separation to keep
statistical uncertainties approximately uniform up to
$t_{\rm s}=1.12$~fm. For larger separations, no such scaling of the statistics is done due to computational cost. This implies that the last two largest time separations have larger uncertainties.
Although these two larger time separations contribute less to the fit process, they remain important for constraining excited-state
contamination. The much larger statistics for the nucleon two-point functions
are inherited from other nucleon-structure projects using the same
ensemble.
\begin{table}[h!]
\centering
\renewcommand{\arraystretch}{1}
\setlength{\tabcolsep}{3pt}

    \begin{tabular}{|l|c|c|c|c|}
    \hline
\multicolumn{5}{|c|}{\textbf{Smearing parameters}} \\
        \hline
        Flavor& $a_G$ & $n_G$ & $a_\text{APE}$ & $n_\text{APE}$\\
        \hline
        Light & 1.0 & 95 & 0.5& 50\\
        Strange & 1.0 & 40 & 0.5& 50\\
        \hline
    \end{tabular}~~
\begin{tabular}{|c|c|c|}
\hline
\multicolumn{3}{|c|}{\textbf{cB211.072.64}} \\
\multicolumn{3}{|c|}{524 configurations} \\
\hline
$t_{\rm s}/a$ & $t_{\rm s}$ [fm] & $n_{src}$ \\
\hline
8 & 0.64 & 1 \\
10 & 0.80 & 3 \\
12 & 0.95 & 9 \\
14 & 1.12 & 27 \\
16 & 1.28 & 27/3 \\
18 & 1.44 & 27/9 \\
\hline
\multicolumn{2}{|l|}{Nucleon 2pt} & 349 \\
\multicolumn{2}{|l|}{$\Lambda$-hyperon 2pt} & 94 \\
\hline
\end{tabular}

\caption{Left: Smearing parameters for the light and strange quark interpolating fields. Right: Statistics for the three-point functions at various source-sink separations, along with the total statistics used for the nucleon and $\Lambda$ two-point functions in the analysis. For the tensor three-point functions, reduced statistics were used at $t_s/$a=16,18, as indicated by the second number in the corresponding entries.}
\label{tab:stat}
\end{table}
\subsection{Ground state extraction}

A controlled treatment of excited-state contamination is essential for a
reliable extraction of ground-state matrix elements. We follow analysis strategies developed in precision studies of nucleon matrix elements \cite{Alexandrou:2024ozj,Alexandrou:2023qbg,Alexandrou:2025vto}. 
Ground-state matrix elements are extracted from ratios of two- and
three-point functions chosen to cancel overlap factors and leading
Euclidean time dependencies. We define the ratio
\begin{equation}
\label{eq:ratio}
R_{\mu}(\vec{p}_N,\vec{p}_\Lambda;t_{\rm s},t_{\rm ins})
\equiv
\frac{
C_{\mu}(\vec{p}_N,\vec{p}_\Lambda;t_{\rm s},t_{\rm ins})
}{
\sqrt{
C_\Lambda(\vec{p}_\Lambda,t_{\rm s})\,
C_N(\vec{p}_N,t_{\rm s})
}
}\,,
\end{equation}
for either scalar- or tensor-current insertion.
In the large-time limit, $\Delta E_B^1 t_{\rm s}\gg 1$, where $\Delta E_B$ is the energy difference between first excited and ground state energy for $B= N, \Lambda$, the ratio becomes
time-independent and yields the desired ground-state matrix elements,
\begin{equation}
\label{eq:LME}
R^{S}_{\mu}(\vec{p}_N,\vec{p}_\Lambda;t_{\rm s},t_{\rm s}/2\pm\tau)
\xrightarrow[\Delta E^1_B t_{\rm s}\gg 1]{}
\Pi^{S}_{\mu}(\vec{p}_N,\vec{p}_\Lambda)\,,
\end{equation}
where $t_{\rm ins}=t_{\rm s}/2\pm\tau$ indicates a plateau region far enough from both source and sink.
The scalar and tensor form factors are obtained from $\Pi$ through
appropriate kinematic projections as described in Sec.~\ref{subsec:decomposition}.

We expand the Euclidean correlators using spectral
decomposition. Retaining the ground state and the first excited state,
the resulting two-point function parametrization is
\begin{equation}
C_{B}(\vec{p},t_{\rm s}) = e^{-t_{\rm s}E_B^{0|\vec{p}|}}(c_B^{0|\vec{p}|}+c_B^{1|\vec{p}|}e^{-t_{\rm s}\Delta E_B^{1|\vec{p}|}}),
\label{two-state}
\end{equation}
where $E_B^{0|\vec{p}|}$ denotes the ground-state energy and
$\Delta E_B^{1|\vec{p}|} = E_B^{1|\vec{p}|} - E_B^{0|\vec{p}|}$ the energy gap
for a given momentum. The amplitudes
\begin{equation}
c_B^{i|\vec{p}|} = \Tr\Big[P_0\langle \chi_B(\vec{p})|B_i(\vec{p})\rangle\langle B_i(\vec{p})|\chi_B(\vec{p})\rangle\,\Big],
\end{equation}
encode the squared overlap of the
interpolator with the corresponding energy eigenstate and, by lattice
rotational symmetry, depend only on $|\vec{p}|$. The trace is over Dirac indices.

Retaining only the ground and first excited state also in spectral decomposition of the three-point functions,
we get
{\thinmuskip=0mu
\medmuskip=1mu
\thickmuskip=2mu
\begin{align}
&C^S_{\mu}(\vec{p}_N,\vec{p}_\Lambda;t_{\rm s},t_{\rm ins})=
e^{-(t_{\rm s}-t_{\rm ins})E_\Lambda^{0|\vec{p}_\Lambda|}-t_{\rm ins}E_N^{0|\vec{p}_N|}}\Big(S^{00}_{\mu}(\vec{p}_N,\vec{p}_\Lambda)\nonumber\\
&~~+S^{01}_{\mu}(\vec{p}_N,\vec{p}_\Lambda)e^{-t_{\rm ins}\Delta E_N^{1|\vec{p}_N|}}+
S^{10}_{\mu}(\vec{p}_N,\vec{p}_\Lambda)e^{-(t_{\rm s}-t_{\rm ins})\Delta E_\Lambda^{1|\vec{p}_\Lambda|}}\nonumber\\
&~~+S^{11}_{\mu}(\vec{p}_N,\vec{p}_\Lambda)e^{-(t_{\rm s}-t_{\rm ins})\Delta E_\Lambda^{1|\vec{p}_\Lambda|}-t_{\rm ins}\Delta E_N^{1|\vec{p}_N|}}\Big)\,,
\end{align}
}
\noindent
where the amplitudes $S^{ij}_{\mu}$ collect the relevant overlap factors
and transition matrix elements,
\begin{multline}
S^{ij}_{\mu}(\vec{p}_N,\vec{p}_\Lambda)=\Tr\Big[P_\mu\langle \chi_\Lambda(\vec{p}_\Lambda)|\Lambda_i(\vec{p}_\Lambda)\rangle\times\\\langle\Lambda_i(\vec{p}_\Lambda)| \bar{u}s|N_j(\vec{p}_N)\rangle\langle N_j(\vec{p}_N)|\chi_N(\vec{p}_N)\rangle \Big].
\end{multline}
An analogous analysis is done for the three-point
functions of the tensor operator. Replacing  the scalar operator by
$\bar{u}\sigma_{\nu\lambda} s$, we obtain for the  amplitudes
\begin{multline}
T^{ij}_{\mu\nu\lambda}(\vec{p}_N,\vec{p}_\Lambda)=\Tr\Big[P_\mu\langle \chi_\Lambda(\vec{p}_\Lambda)|\Lambda_i(\vec{p}_\Lambda)\rangle\times\\\langle\Lambda_i(\vec{p}_\Lambda)| \bar{u}\sigma_{\nu\lambda} s|N_j(\vec{p}_N)\rangle\langle N_j(\vec{p}_N)|\chi_N(\vec{p}_N)\rangle \Big].
\end{multline}

The target ground-state matrix elements, follow from the leading amplitude via
\begin{equation}
    \Pi^\text{S}_{\mu}(\vec{p}_N,\vec{p}_\Lambda) = \frac{S^{00}_{\mu}(\vec{p}_N,\vec{p}_\Lambda)}{\sqrt{c^0_\Lambda(\vec{p}_\Lambda)c^0_N(\vec{p}_N)}}\,,
\end{equation}
for the scalar, and
\begin{equation}
    \Pi^\text{T}_{\mu\nu\lambda}(\vec{p}_N,\vec{p}_\Lambda) = \frac{T^{00}_{\mu\nu\lambda}(\vec{p}_N,\vec{p}_\Lambda)}{\sqrt{c^0_\Lambda(\vec{p}_\Lambda)c^0_N(\vec{p}_N)}}\,,
\end{equation}
for the tensor.

\begin{figure*}
\centering
{\includegraphics[width =\linewidth]{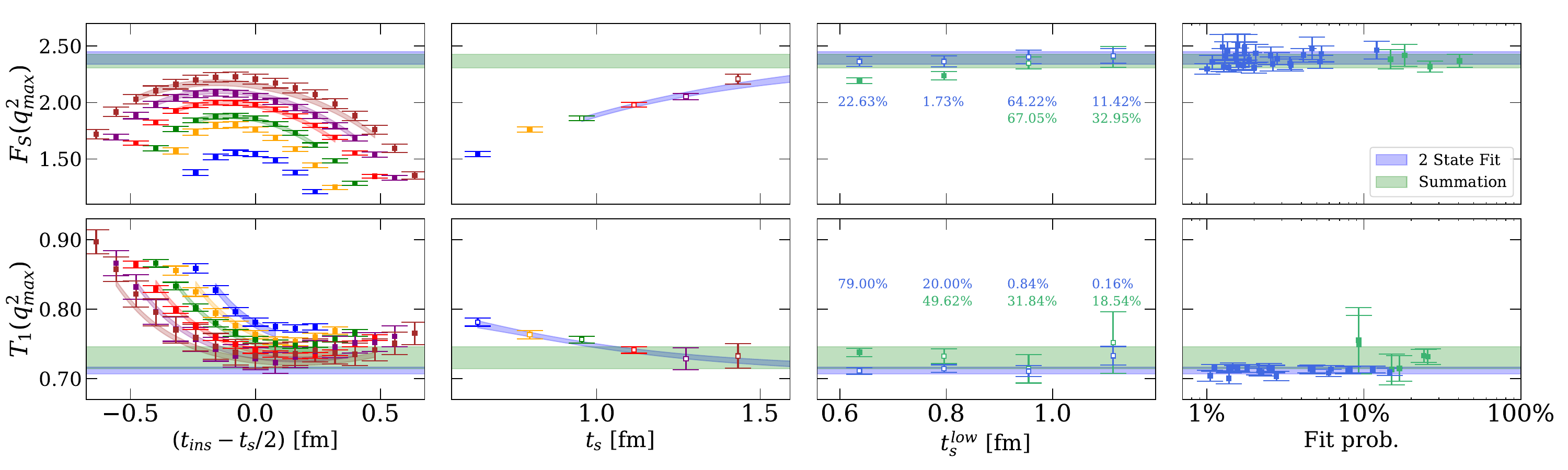}}

\caption{Results for the ratios of three- to two-point correlation functions, as defined in Eq.~\eqref{eq:ratio}, are shown at the kinematic point $\vec{p}_N=\vec{p}_\Lambda=\vec{0}$, corresponding to $q^2=q^2_{\rm max}$. After appropriate normalization and renormalization, these ratios, in the large time limit, yield the scalar and tensor form factors. The leftmost panels display the dependence on the insertion time $t_{\rm ins}$ for the different source--sink time separations $t_{\rm s}$, while the second-column panels show the $t_{\rm s}$ dependence of the midpoint values at $t_{\rm ins}=t_{\rm s}/2$. Colors are matched across panels to identify the same $t_{\rm s}$ values. The third-column panels present the ground-state matrix elements extracted from the two-state (blue) and summation method (green) analysis, as a function of
the lowest value of $t_s$ included in the fit, $t_s^{low}$. The filled points are excluded from the model-averaging procedure, as they correspond to fit ranges that do not exhibit stable convergence. The rightmost panels show the fitting probability larger than $1\%$, characterizing the ground-state matrix elements coming from the various fits. The green bands correspond to the model-averaged results obtained using the summation method, and the blue bands to the model-averaged results from the two-state fits. For improved visualization, the data have been rescaled by $e^{-(m_\Lambda - m_N)(t_{\rm ins} - t_{\rm s}/2)}$ to remove the residual time dependence associated with the ground-state energy difference. 
}
\label{fig:fitFFs}
\end{figure*}

\subsection{Fitting strategy}

We determine the ground-state
matrix elements using two complementary approaches: the summation method, where only the ground state is retained and two-state fits.

In the \emph{summation method}, we form an improved ratio designed to
remove the residual time dependence associated with the operator
insertion time $t_{\rm ins}$ \cite{Alexandrou:2013joa,Alexandrou:2011db,Alexandrou:2006ru}.
Summing over insertion times in the interval $[n,t_{\rm s}-n]$ leads to
\begin{multline}
\label{eq:summeth}
S^{\rm S}_{\mu}
(\vec{p}_N,\vec{p}_\Lambda;t_{\rm s})\equiv \sum_{t_{\rm ins}=n}^{t_{\rm s}-n}R^{\rm S}_{\mu}
(\vec{p}_N,\vec{p}_\Lambda;t_{\rm s},t_{\rm ins})\times\\ \frac{\sqrt{C_\Lambda(\vec{p}_\Lambda;t_{\rm ins})
 C_N(\vec{p}_N;t_{\rm s}-t_{\rm ins})}}{\sqrt{C_N(\vec{p}_N;t_{\rm ins})C_\Lambda(\vec{p}_\Lambda,t_{\rm s}-t_{\rm ins})
}},
\end{multline}
which generates a geometric series in the excited-state contributions
\cite{Maiani:1987by,Capitani:2012gj}. The summed ratio behaves as
\begin{equation}
S^{\rm S}_{\mu}
(\vec{p}_N,\vec{p}_\Lambda;t_{\rm s})= c + \Pi_{\mu}^{\rm S}(\vec{p}_N,\vec{p}_\Lambda)\, t_{s} + \dots    
\end{equation}
so that the desired matrix element is obtained from the slope of a
linear fit in $t_{\rm s}$. The omitted terms are dominated by
excited-state contamination and are exponentially suppressed with
$t_{\rm s}$ instead of $t_s-t_{\rm ins}$. In practice, the method depends on the summation range
parameter $n$ and the minimum source--sink time separation $t_{\rm s,min}$
used in the linear fit. To reduce sensitivity to these analysis choices,
we perform fits over multiple admissible windows and remove the residual
dependence through model averaging based on the Akaike Information
Criterion (AIC) \cite{Jay:2020jkz,Neil:2022joj}. For each fit $i$, we
assign a weight $w_i$ given by
\begin{equation}\label{eq:weight}
\log(w_i) = -\frac{\chi_i^2}{2} + N_{\text{dof}, i},
\end{equation}
where $N_{\text{dof}, i}=N_{\text{data}, i}-N_{\text{params}, i}$ and
$\chi_i^2$ is obtained from the correlated fit.

 For the two-point functions, we
fit the effective energy 
\begin{equation}
    \label{eq:fit1}
E_B^{\rm eff}(t_{\rm s},\vec{p}) = \log\frac{C_B(t_{\rm s},\vec{p})}{C_B(t_{\rm s}+1,\vec{p})}= E^B_0(\vec{p})+\dots,
\end{equation}
and perform two-state fits using Eq.~(\ref{two-state}). In the \emph{two-state fit} approach, 
we fit  the ratio in Eq.~\eqref{eq:ratio}, which approaches
the matrix element in the large-time limit as summarized in
Eq.~\eqref{eq:LME}. To enhance the determination of the $\Lambda$--$N$
mass splitting, we also include the ratio
\begin{equation}
\label{eq:fit2m}
m_{\Lambda/N}^{\rm eff}(t_{\rm s}) = \log\frac{C^{\Lambda}(t_{\rm s})C^{N}(t_{\rm s}+1)}{C^{N}(t_{\rm s})C^{\Lambda}(t_{\rm s}+1)}=m_\Lambda-m_N+\dots,
\end{equation}
which isolates $m_\Lambda-m_N$ asymptotically. Since the energies enter both two- and three-point functions, we perform
correlated combined fits at each value of the momentum transfer.

More specifically, for each form factor at $q^2=q^2_{\max}$ we carry out
a simultaneous fit to
$E_N^{\rm eff}(t_{\rm s},\vec{0})$,
$E_\Lambda^{\rm eff}(t_{\rm s},\vec{0})$,
$m_{\Lambda/N}^{\rm eff}(t_{\rm s})$,
and the ratio defined in Eq.~\eqref{eq:ratio}.
In this fit, the parameters
$E_B^{0|\vec{0}|}$,
$c_B^{0|\vec{0}|}$,
$c_B^{1|\vec{0}|}$,
and $\Delta E_B^{1|\vec{0}|}$
are shared.

For $q^2 \neq q^2_{\max}$, we perform a correlated combined fit to
$E_N^{\rm eff}(t_{\rm s},\vec{p})$
and the ratio in Eq.~\eqref{eq:ratio}.
In this case, the nucleon ground-state energy is constrained using the dispersion relation $E_N^{0|\vec{p}|} = \sqrt{(E_N^{0|\vec{0}|})^2 + |\vec{p}\,|^2}$,
while the $\Lambda$-baryon ground-state parameters
$E_\Lambda^{0|\vec{0}|}$,
$c_\Lambda^{1|\vec{0}|}$ and
$\Delta E_B^{1|\vec{0}|}$
are fixed from the zero-momentum analysis and used in the denominator
of Eq.~\eqref{eq:ratio}. The nucleon first excited state $\Delta E_N^{1|\vec{p}_N|}$ is shared between the two- and three-point functions, while the remaining parameters are left free in the fit. 
$T_2$, $T_3$, and $T_4$ are fitted simultaneously, sharing the excited-state parameters in the three-point functions, while their zero-momentum parameters are adopted from the fit of $T_1$.
The associated fit windows are controlled by a
set of hyperparameters ($t_{{\rm s,min,\Lambda}}$, $t_{{\rm s,min,N}}$,
$t_{{\rm s,min,3pt}}$, $t_{{\rm ins,min,3pt}}$, and
$t_{{\rm ins,max,3pt}}$), which are varied within reasonable ranges. The remaining dependence on these choices is
removed by AIC-based model averaging with weights defined in
Eq.~\eqref{eq:weight}. Similarly, for the summation method the associated fit windows are controlled by the n-parameter appearing in Eq. \ref{eq:summeth}.

\begin{figure*}
\centering
{\includegraphics[width = \linewidth]{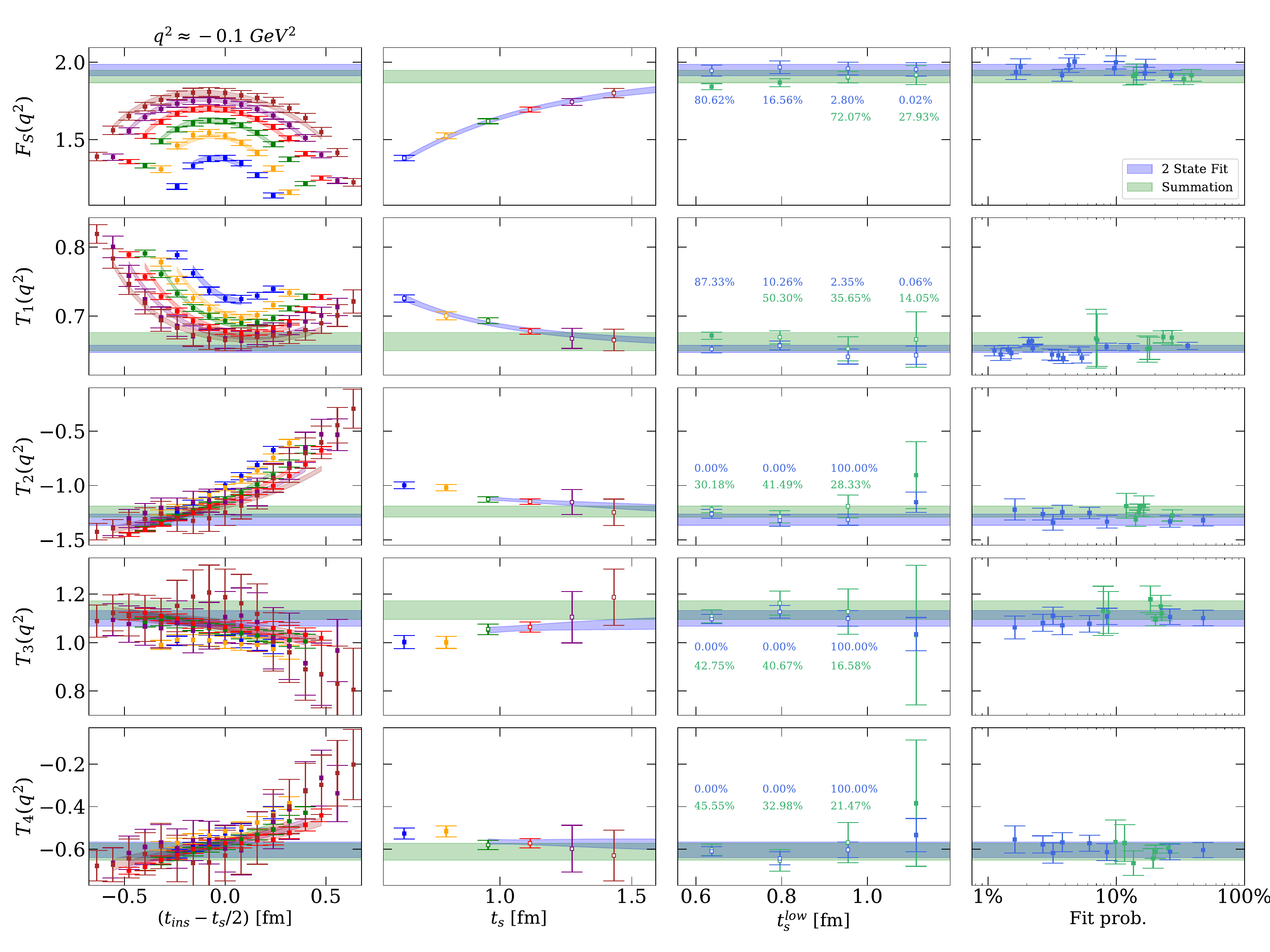}}
\vspace{-0.3cm}
\caption{Results for the ratios of three- to two-point correlation functions, similar to Fig. \ref{fig:fitFFs}, with $q^2\approx-0.1$ GeV$^2$.
}
\label{fig:fitFFs2}
\end{figure*}
The effective mass analysis has already been presented in Fig. 7 of \cite{Bacchio:2025auj}, while in Fig.~\ref{fig:fitFFs} the fits for the scalar and tensor matrix elements are presented at $q^2_{\max}$, corresponding to
$\vec{p}_N=\vec{p}_\Lambda=\vec{0}$, and Fig. \ref{fig:fitFFs2} corresponds to $q^2\approx-0.1$ GeV$^2$. The analysis is repeated for all
accessible nucleon momenta $\vec{p}_N$, while keeping
$\vec{p}_\Lambda=\vec{0}$ fixed for the three-point functions in the fixed sink setup.

\subsection{Decomposition of matrix elements}\label{subsec:decomposition}

The ground-state matrix elements are converted into form factors by
solving a linear system of kinematic relations in the helicity basis.
For $\vec{p}_\Lambda=\vec{0}$, the non-vanishing components relevant for
the scalar current takes the form
\begin{equation}
\Pi^{S}_{0}(\vec p_N)
= {\cal K}\,(E_N+m_N)\,F_S(q^2)\,,
\end{equation}
and for tensor currents
{\thinmuskip=0mu
\medmuskip=1mu
\thickmuskip=2mu
\begin{align}
\Pi^{T}_{ijk}(\vec p_N)
&= {\cal K}\,\epsilon_{jkl}\,p_N^{l}p_N^{i}\,
\frac{T_3(q^2)}{m_N}\,(1-|\epsilon_{ijk}|)  \\
\quad
+{\cal K}\,\epsilon_{ijk}
&\left[
(E_N+m_N)\,T_1(q^2)
-\frac{(p_N^j)^2+(p_N^k)^2}{m_N}\,T_3(q^2)
\right]\nonumber
\end{align}
\begin{equation}
    \Pi^{T}_{i0j}(\vec p_N)
= -i\,{\cal K}\,\epsilon_{ijk}\,p_N^{k}
\left[
T_2(q^2)-T_1(q^2)+\frac{E_N}{m_N}T_3(q^2)
\right]
\end{equation}
\begin{equation}
    \Pi^{T}_{00i}(\vec p_N)
= {\cal K}\,p_N^{i}
\left[
T_1(q^2)-T_2(q^2)+T_3(q^2)
+\frac{m_N+E_N}{m_N}T_4(q^2)
\right]
\end{equation}
}
\noindent
with kinematic prefactor
\begin{equation}
{\cal K}=\left[2E_N(E_N+m_N)\right]^{-1/2}.
\end{equation}

For this transition, the system has rank $1^{\text{S}}+4^{\text{T}}$,
matching the number of independent form factors necessary for the full decomposition of matrix elements in Eqs.~(\ref{eq:scalar_decomp},\ref{eq:tensor_decomp}). An exception exists in the tensor form factors when two of the three momentum components are zero. In that case, we have 4 unknowns with three equations and therefore the determination of the form factors becomes impossible. For this reason, these kinematic points are excluded
from the tensor analysis. After averaging
linearly dependent matrix elements, the system becomes exactly
solvable. In our workflow, the form factor decomposition is carried out
at each value of $q^2$ prior to fitting the $q^2$ dependence. This is
justified because the decomposition involves only linear combinations of
matrix elements and therefore preserves the spectral structure assumed
in the summation and two-state fit Ans\"atze.

\subsection{The $z$-expansion}\label{z:exp}

The momentum-transfer dependence of each form factor is described using
the model-independent $z$-expansion \cite{Hill:2010yb},
\begin{align}
S(q^2)=\sum_{k=0}^{k_{\max}}a_k z^k~\text{with}~
z=\frac{\sqrt{t_{\rm cut}-q^2}-\sqrt{t_{\rm cut}}}{\sqrt{t_{\rm cut}-q^2}+\sqrt{t_{\rm cut}}},
\end{align}
where the branch point is set by
$t_{\rm cut}=(m_\pi+m_K)^2$. Following standard practice~\cite{Hill:2010yb}, we enforce a vanishing behavior in the limit $q^2\to-\infty$ by setting
$a_{k_{\max}}=-\sum_{k=0}^{k_{\max}-1}a_k$. The corresponding charges and
radii are obtained from the expansion coefficients as
\begin{equation}
\label{eq:ch_rad}
f_{S}\equiv F_{S}(0)=a_0\quad\text{and}\quad
\langle r^2_{F_{S}} \rangle = -\frac{3a_1}{2a_0 t_{\rm cut}},
\end{equation}
which are related to the leading low-$q^2$ expansion as
\begin{equation}
    F_i(q^2) = f_i \left(1+\frac{\langle r^2_{F_i} \rangle}{6} q^2\right) + \mathcal{O}(q^4)\,.\label{eq:lowq2}
\end{equation}
The $q^2$ dependence is fitted with $k_{\max}=3$ for all form factors, and the constraint $a_3=-(a_0+a_1+a_2)$ is used to enforce vanishing behavior as $q^2\to -\infty$.

\begin{figure*}
\includegraphics[width = 5.5in]{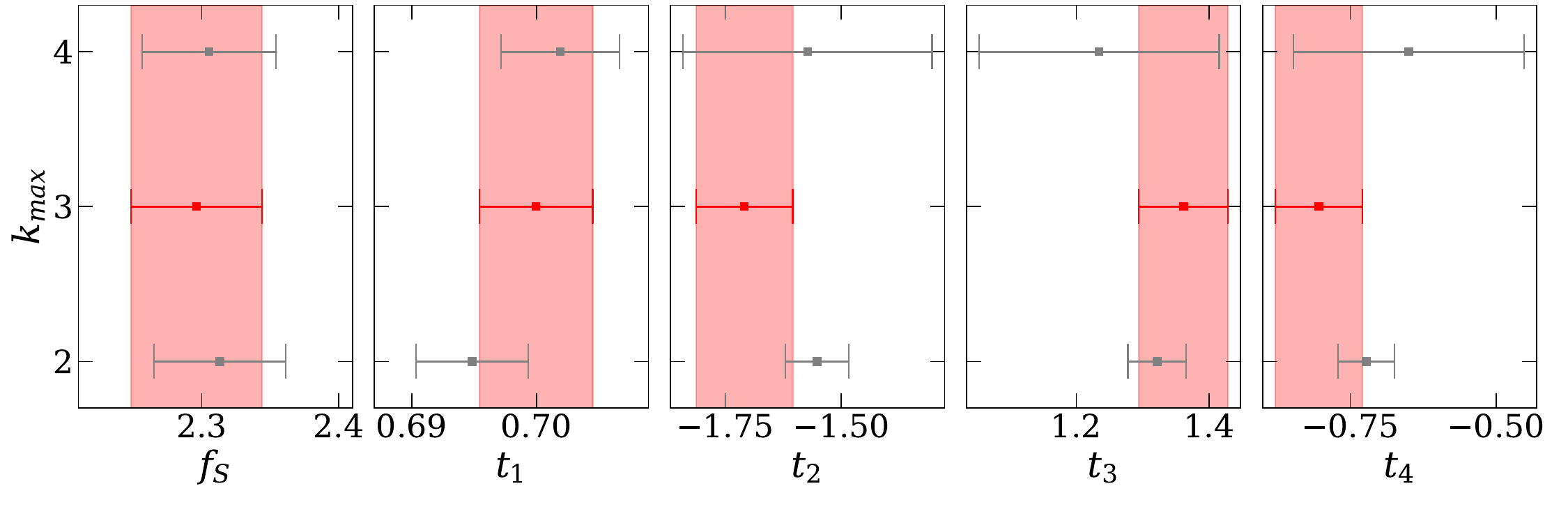}
\includegraphics[width = 5.5in]{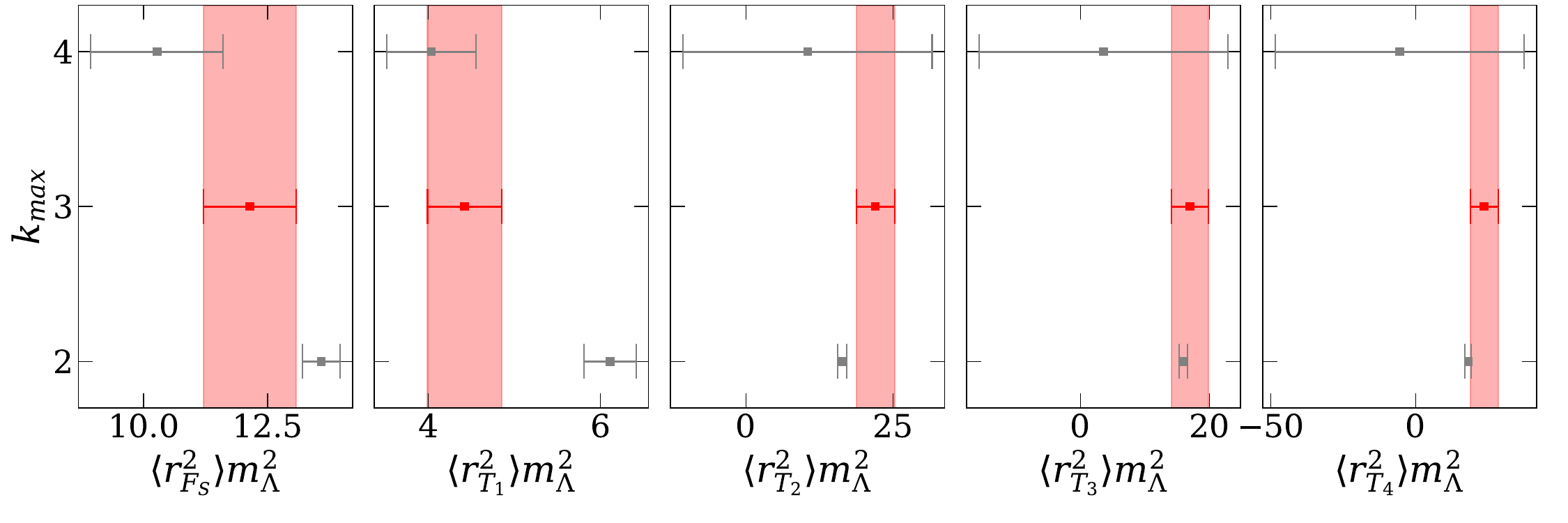}
\caption{
Stability of the $z$-expansion fits under changes of the order
$k_{\max}$. The upper panels show the scalar and tensor couplings
$f_S$ and $t_i \equiv T_i(0)$, while the lower panels show the corresponding
radii multiplied by $m_\Lambda^2$. Results are shown for $k_{\max}=2,3,$ and $4$.
The red points and bands indicate the choice, $k_{\max}=3$, used in the
final analysis. 
}
\label{fig:stab}
\end{figure*}

The choice $k_{\max}$ is motivated by the stability study shown in
Fig.~\ref{fig:stab}. The  analysis is performed with $k_{\max}=2,3,$ and $4$ giving
compatible results for all couplings and radii. We therefore use
$k_{\max}=3$ as our final  choice, since increasing the truncation order does not
lead to a statistically significant change in the extracted quantities.

\begin{figure*}
\centering
{\includegraphics[width = 0.75\linewidth]{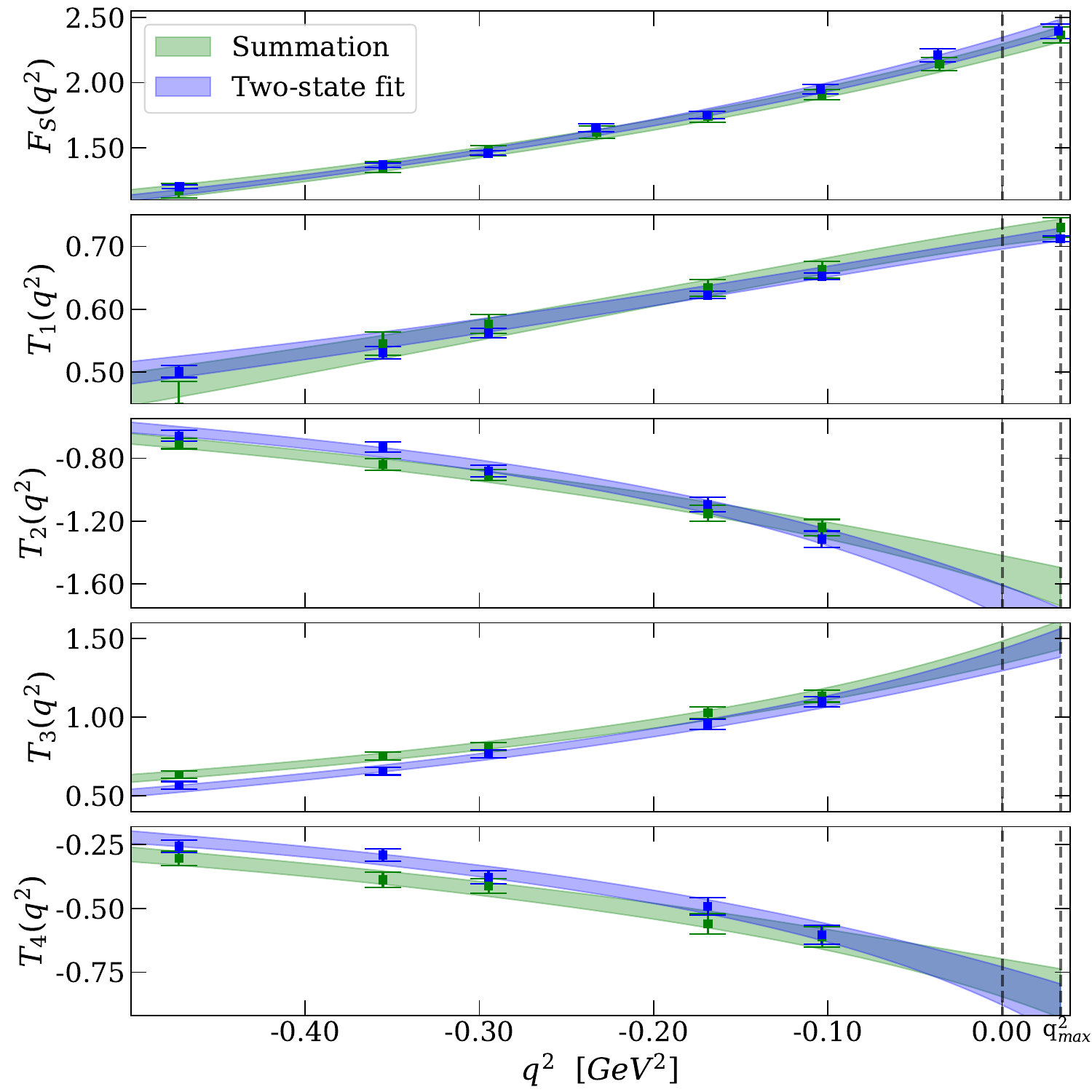}}

\caption{Momentum-transfer dependence of the scalar and tensor form factors. Blue markers denote results obtained from  two-state fits, while green markers correspond to the summation method. The associated $z$-expansion descriptions are shown as blue and green bands, respectively. The vertical dashed lines delimit the physical region relevant for the semileptonic decay, $q^2 \in [0, q^2_{\rm max}]$.}
\label{fig:ffs}
\end{figure*}

\section{Results}\label{sec:results}

\subsection{Scalar and tensor form factors}
\label{subsec:results_STff}

Fig.~\ref{fig:ffs} displays our lattice QCD determination of the scalar and
tensor $\Lambda\to p$ form factors as a function of the momentum transfer $q^2$. The discrete points correspond to the ground-state matrix elements extracted at the available lattice kinematics, while the shaded bands show the correlated $z$-expansion fits used to describe the full $q^2$-dependence in the spacelike region accessible on the lattice and to interpolate to the physical timelike region relevant for the decay, $q^2\in[0,q^2_{\rm max}]$. We observe good agreement between the summation and two-state fit analyses for all form factors, but in order to ensure a controlled treatment of excited-state effects, we adopt the two-state fit results. 
The resulting charges and the radii extracted by fitting the form factors are given in Table~\ref{tab:chargeradi}. The fit parameters of the $z$-expansion  are given in Appendix \ref{app:pars}.

\begin{table}
    \centering
    \begin{tabular}{ccc}
    \hline
    \hline
        $f_S$ & ~~$\langle r^2_{F_S} \rangle m_\Lambda^2$~~ & corr.\\
        \hline
        2.296(48)& 12.14(94)&0.43\\
                \hline
                \hline
                \vspace{2mm}
    \end{tabular}
    \\
\begin{tabular}{cccc}
\hline
\hline
        $t_1$  & $t_2$ & $t_3$ & $t_4$  \\
        0.7000(46) & -1.71(10) & 1.361(67) & -0.804(75)  \\
                \hline

        $\langle r^2_{T_1} \rangle m_\Lambda^2$ & $\langle r^2_{T_2} \rangle m_\Lambda^2$& $\langle r^2_{T_3} \rangle m_\Lambda^2$& $\langle r^2_{T_4} \rangle m_\Lambda^2$ \\
         4.42(43) & 22.0(3.2) & 17.0(2.9)& 23.6(4.8) \\  
        \hline
        \hline
        \vspace{0.5mm}
    \end{tabular}

\begin{blockarray}{ccccccccc}
~\text{corr.}~~ & $t_1$ & $\langle r^2_{T_1} \rangle$ & $t_2$ & $\langle r^2_{T_2} \rangle$ & $t_3$ & $\langle r^2_{T_3} \rangle$& $t_4$ & $\langle r^2_{T_4} \rangle$ \\
\begin{block}{c(cccccccc)}
$t_1$&1.00 & -0.05 & -0.19 & 0.04 & 0.09 & -0.00 & -0.14 & 0.02  \\
$\langle r^2_{T_1} \rangle$&-0.05 & 1.00 & 0.04 & 0.06 & -0.07 & 0.03 & 0.07 & 0.04 \\
 $t_2$&-0.19 & 0.04 & 1.00 & -0.79 & -0.85 & -0.71 & 0.90 & -0.70  \\
$\langle r^2_{T_2} \rangle$&0.04 & 0.06 & -0.79 & 1.00 & 0.65 & 0.84 & -0.69 & 0.88\\
$t_3$ &0.09 & -0.07 & -0.85 & 0.65 & 1.00 & 0.82 & -0.94 & 0.72\\
$\langle r^2_{T_3} \rangle$&0.00 & 0.03 & -0.71 & 0.84 & 0.82 & 1.00 & -0.77 & 0.93\\
$t_4$ &-0.14 & 0.07 & 0.90 & -0.69 & -0.94 & -0.77 & 1.00 & -0.78\\
$\langle r^2_{T_4} \rangle$ &0.02 & 0.04 & -0.70 & 0.88 & 0.72 & 0.93 & -0.78 & 1.00 \\
\end{block}\\
\end{blockarray}
    
    \caption{\label{tab:chargeradi}Charges and radii of the scalar and tensor form factors and the corresponding correlation matrix. The radii are provided in units of the experimental value of $m_\Lambda^2$ with $m_\Lambda = 1.11568$\,GeV.}
\end{table}

\subsection{Comparison of scalar and tensor charges}
\label{subsec:results_STcharges}

\begin{figure}
\centering
\includegraphics[width = 0.9\linewidth]{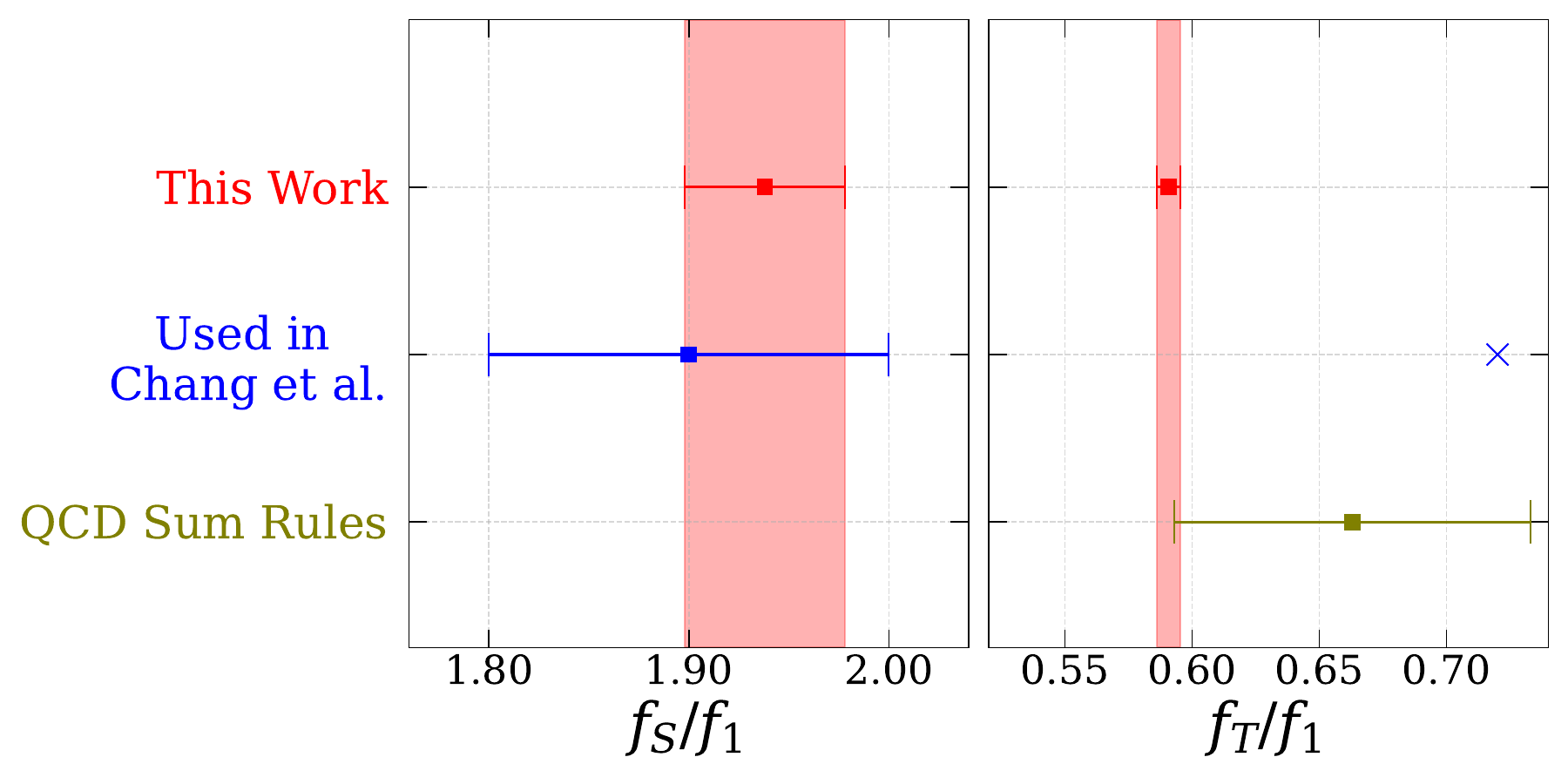}
\caption{Comparison of the scalar and tensor charge ratios $f_S/f_1$ and $f_T/f_1$ obtained in this work with other existing results. The phenomenological inputs adopted in Ref.~\cite{Chang:2014iba} for $f_S/f_1$ was inferred from the conserved vector current relation $f_S/f_1=\Delta/(m_s-m_u)$~\cite{Gonzalez-Alonso:2013ura}, while $f_T/f_1$ was estimated using model calculations~\cite{Ledwig:2010tu}.For $f_T/f_1$, we additionally compare with the QCD sum-rule determination of Ref.~\cite{Zhang:2024ick}.}
\label{fig:ST_FFs}
\end{figure}

In Fig.~\ref{fig:ST_FFs}, we compare the charge ratios $f_S/f_1$ and $f_T/f_1$ obtained in this work, with other results obtained from
phenomenological analysis~\cite{Chang:2014iba} or QCD sum-rule results ~\cite{Zhang:2024ick}. 
Our values are
\begin{equation}
    \frac{f_S}{f_1}{=}1.938(40)\qq{and} \frac{f_T}{f_1}{=}0.5908(47),
\end{equation}
with $f_1 = 1.1849(57)$  determined in Ref.~\cite{Bacchio:2025auj}.
Compared to the phenomenological estimates used in Ref.~\cite{Chang:2014iba}, the lattice determination yields a compatible value with reduced uncertainty in the
scalar channel, while for the tensor ratio we observe a clear difference relative to their estimate. 

\subsection{Decay rates, ratio, and sensitivity}
\label{subsec:results_decays}

\begin{figure}
\centering
{\includegraphics[width = \linewidth]{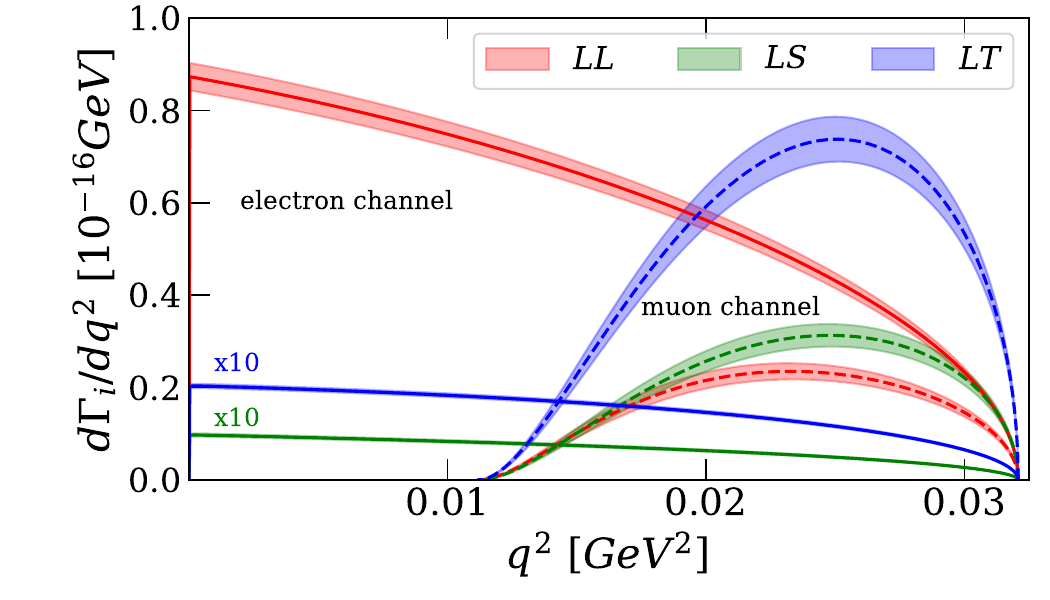}}
\caption{Differential decay rates for the electron and muon channels as a function of $q^2$. The term $LL$ denotes the Standard Model left-handed $V{-}A$ contribution, while $LS$ and $LT$ represent non-standard scalar and tensor interference terms with the SM amplitude. The latter decay rates are then multiplied by $\epsilon_S$ and $\epsilon_T$ respectively as in Eq.~\eqref{eqn:dif}. For visualization purposes, the electron channel contributions by $LS$ and $LT$ are multiplied by a factor of 10. The  inputs are $m_e = 0.511\,\mathrm{MeV}$, $m_\mu = 0.10566\,\mathrm{GeV}$, $G_F = 1.16638 \times 10^{-5}(1 + 0.0105)\,\mathrm{GeV}^{-2}$~\cite{ParticleDataGroup:2024cfk}, where the multiplicative factor accounts for radiative corrections~\cite{Garcia:1985xz}, and $| V_{us}| = 0.22431(85)$~\cite{ParticleDataGroup:2024cfk}. The baryon masses entering our quoted results are determined from the same lattice QCD ensemble as the form factors,
$m_\Lambda = 1.1263(23)$\,GeV, $m_N = 0.9471(28)$\,GeV, and
$m_\Lambda - m_N = 0.1792(24)$\,GeV. This choice keeps the
kinematic input consistent with the finite-lattice-spacing calculation.}
\label{fig:dec}
\end{figure}

\begin{figure}
\centering
{\includegraphics[width = \linewidth]{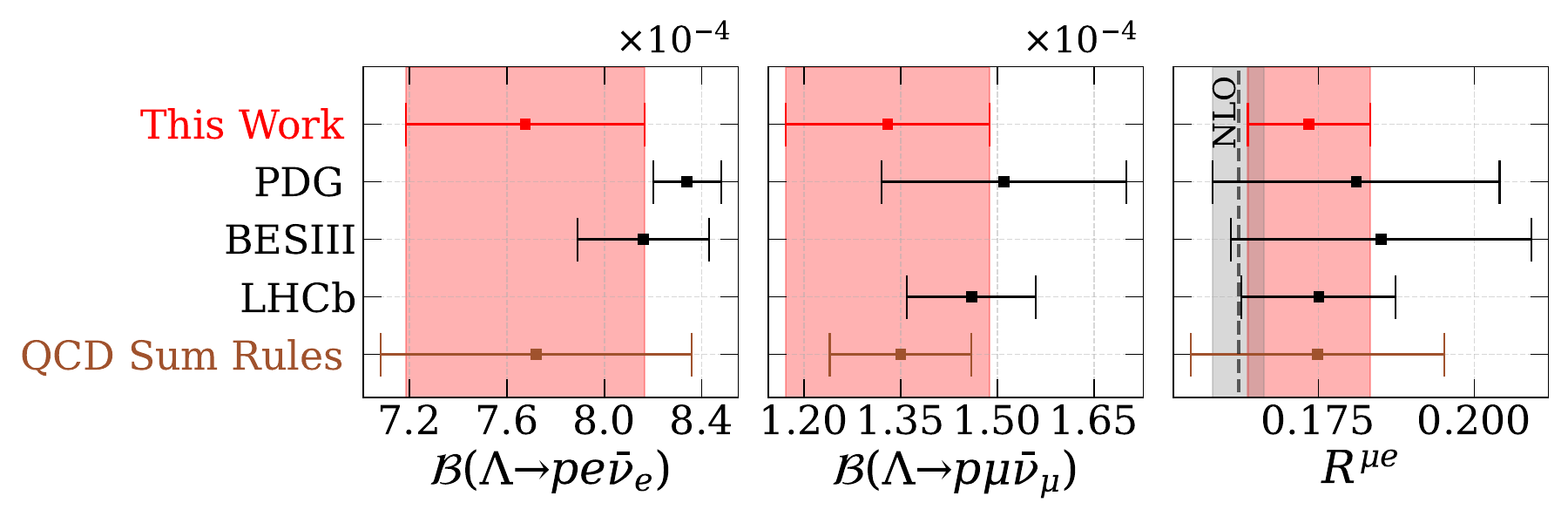}} 
\caption{Results on the branching fractions in the electron and muon channels and their ratio. Our results are shown in red, and are obtained using the nucleon and $\Lambda$
masses determined from lattice QCD. The experimental values and their ratios, $R^{\mu e}_{\rm PDG}$ =  0.181(23), $R^{\mu e}_{\rm PDG/BESIII}$ =  0.185(24) and $R^{\mu e}_{\rm LHCb/PDG}$ =  0.175(12), are shown by the black points \cite{ParticleDataGroup:2024cfk,BESIII:2025hgj,LHCb:2025wld}, and the QCD sum rules value by the brown points~\cite{Zhang:2024ick,Zhang:2025qmg}. The gray dashed line represents the perturbative value of the ratio from Eq.~\eqref{eq.Rat}, which is $R^{\mu e}_{pert.}=0.1622(41)$ when the experimental masses~\cite{ParticleDataGroup:2024cfk} are adopted. The associated error is $\delta^2\approx 2.5\%$. 
}
\label{fig:decrarte}
\end{figure}

Fig.~\ref{fig:dec} shows the differential decay rate via standard and non-standard interactions as a function of the momentum transfer $q^2$. We note that scalar and tensor contributions have to be multiplied by $\epsilon_S$ and $\epsilon_T$ respectively as in Eq.~\eqref{eqn:dif}.
After integrating the differential decay rates over $q^2$ as in Eq.~\eqref{eq:decay_int}, we obtain the non-perturbative value of the sensitivity coefficients defined in Eq.~\eqref{eq:Rlin} with
\begin{equation}\label{eq:final_r}
r_S = 1.271(29)\qq{and} r_T = 2.975(34)\,.
\end{equation}
Using, instead, the perturbative Eq.~\eqref{Eq:sens} we would obtain
\begin{equation}\label{eq:final_r}
r_S^{pert.} = 1.312(81)\qq{and} r_T^{pert.} = 3.31(17)\,,
\end{equation}
which shows a few percent relative deviation compared to their non-perturbative estimation, namely
\begin{equation}
    \Delta{r_S}=-3.1(5.5)\% \qq{and}\Delta{r_T}=-10.2(53)\%\,,
\end{equation}
with $\Delta r=(r-r^{pert.})/r^{pert.}$.

In Fig.~\ref{fig:decrarte}, we present an updated comparison of the Standard Model decay rates. Relative to our previous work in Ref.~\cite{Bacchio:2025auj}, we now include recent BESIII results for the electron mode~\cite{BESIII:2025hgj} and LHCb results for the muon mode~\cite{LHCb:2025wld}. Our lattice QCD
result, shown by the red point, is obtained using the baryon masses determined
from the same lattice ensemble as the form factors,
$m_\Lambda = 1.1263(23)$\,GeV, $m_N = 0.9471(28)$\,GeV, and
$m_\Lambda-m_N = 0.1792(24)$\,GeV. This choice keeps the kinematic input
consistent with the finite-lattice-spacing calculation. For the ratio of the muon-to-electron modes, we additionally compare with the perturbative estimate from Eq.~\eqref{eq.Rat}, $R^{\mu e}_{pert.}=0.1622(41)$, which is independent of hadron form factors. Relative to this value, we find a positive deviation of up to $\Delta{R^{\mu e}}=7.0(6.6)\%$ when using the lattice baryon masses, a value consistent with other experimental determinations and QCD sum-rule estimates, as well as with the expected $\delta^2\approx2.5\%$ effects in the perturbative determination.

\subsection{Constraints to $\epsilon_S$ and $\epsilon_T$} \label{subsec:results_contours}

\begin{table}[t]
\centering

\begin{tabular}{lc||ccc|c}
\hline\hline
& \textbf{This work} & \multicolumn{4}{c}{Using external inputs and expansions}\\
& $\Lambda \to p$ & $\Sigma^- \to n$ & $\Xi^0 \to \Sigma^+$ & $\Xi^- \to \Lambda$ & Ref. \\
\hline
$g_1/f_1$        & 0.6902(44) & $-0.340(17)$ & 1.220(5) & 0.250(5) & \cite{ParticleDataGroup:2024cfk}\\
$f_S/f_1$        & 1.938(40)  & 2.80(14)     & 1.36(7)   & 2.25(11) & \cite{Chang:2014iba}\\
$f_T/f_1$        &0.5908(47)   & $-0.284(46)$      & 1.28(11)     & 0.288(51)  &   \cite{Zhang:2024ick} \\
\hline
$r_S$                  & 1.271(29)      & 2.79(21)         & 0.469(48)    & 2.84(27)   &  \eqref{Eq:sens}  \\
$r_T$                  & 2.975(34)     &1.15(20)         &  6.45(69)      & 1.09(25)   &  \eqref{Eq:sens}\\
\hline
$R^{\mu e}_{\rm SM}$                  & 0.1735(98)      & 0.446(21)         & 0.00841(26)    & 0.2729(66)   &  \eqref{eq.Rat} \\
$R^{\mu e}_{\rm exp}$                  & 0.175(12)      & 0.442(42)          & 0.0092(14)    & 0.74(51)   &  \cite{ParticleDataGroup:2024cfk} \\
\hline\hline
\end{tabular}
\caption{
Inputs and results used to constrain non-standard interactions in semileptonic hyperon decays. For $\Lambda\to p$, $r_{S,T}$ and $R^{\mu e}_{\rm SM}$ are determined nonperturbatively from our lattice-QCD form factors, while for $\Sigma^- \to n$, $\Xi^0 \to \Sigma^+$, and $\Xi^- \to \Lambda$ we use phenomenological and experimental inputs together with perturbative expansions (see main text and provided references). The table also lists the experimental ratios $R^{\mu e}_{\rm exp}$ (PDG averages, except for $\Lambda\to p$, where the muon mode uses the recent LHCb result~\cite{LHCb:2025wld}).}
\label{tab:rsrt}
\end{table}

\begin{figure}
\centering
{\includegraphics[width = 0.9\linewidth]{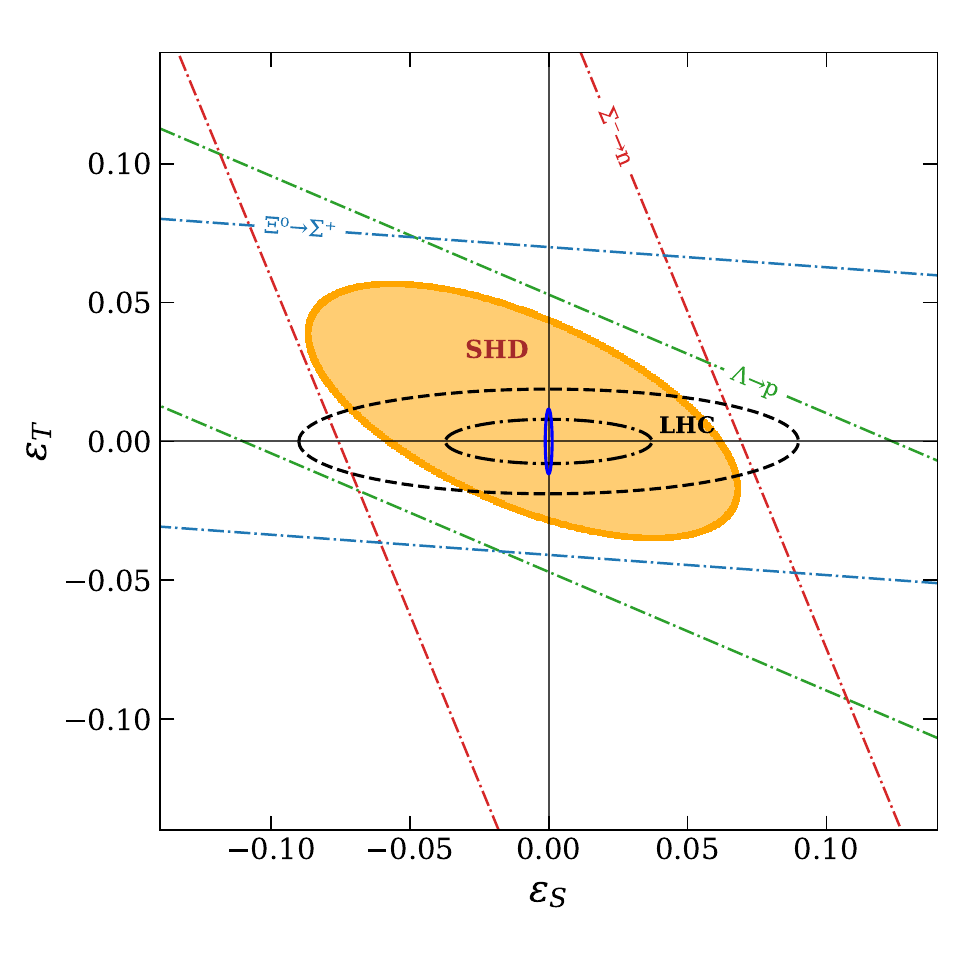}}
\caption{$90\%$ C.L.\ constraints on the scalar and tensor  coefficients $(\epsilon_S,\epsilon_T)$ at $\mu=2$~GeV probing new physics. The filled orange ellipse shows the semileptonic hyperon decays constraint obtained in the present work. Black dashed lines denote bounds from high-$p_T$ LHC searches for non-standard $us\to e\nu$ interactions in $pp\to e+\slashed{E}_T+X$~\cite{Cirigliano:2012ab,Chang:2014iba}, with the outer line corresponding to $\sqrt{s}=7$~TeV and the inner to $8$~TeV. The blue contour shows the constraint from semileptonic kaon decays ($K_{\ell3}$) from Ref.~\cite{Gonzalez-Alonso:2016etj}.}
\label{fig:cont}
\end{figure}

Having determined the semileptonic decay rates from the form factors for of $\Lambda \to p$, we can use it to provide constraints on non-standard interactions by combining inputs from  hyperon $s\to u$ channels, following Ref.~\cite{Chang:2014iba}. For $\Lambda\to p$, our lattice QCD form factors provide the most precise  input and are used to determine nonperturbatively both the sensitivities $r_{S,T}$ and the muon-to-electron mode ratio $R^{\mu e}_{\rm SM}$. For $\Sigma^- \to n$, $\Xi^0 \to \Sigma^+$, and $\Xi^- \to \Lambda$, lattice QCD results for scalar and tensor form factors are not available, and we instead rely on phenomenological or experimental inputs together with perturbative expansions. In particular, $g_1/f_1$ is taken from the PDG average of experimental measurements~\cite{ParticleDataGroup:2024cfk}, while $f_S/f_1$ is estimated phenomenologically following Ref.~\cite{Chang:2014iba} using the conserved vector current relation $f_S/f_1=\Delta/(m_s-m_u)$~\cite{Gonzalez-Alonso:2013ura}. Results for $f_T/f_1$ are taken from QCD sum rules~\cite{Zhang:2024ick}. The sensitivities $r_S$ and $r_T$ are determined using Eq.~\eqref{Eq:sens}, with the dominant uncertainties arising from those of the above couplings. The ratio $R^{\mu e}_{\rm SM}$ is computed from Eq.~\eqref{eq.Rat}; here the leading uncertainty originates from $\delta^2$ effects, since the ratio is independent of hadronic contributions. The resulting values are summarized in Table~\ref{tab:rsrt}, together with the experimental muon-to-electron decay-mode ratio $R^{\mu e}_{\rm exp}$. For all modes we use the PDG averages~\cite{ParticleDataGroup:2024cfk}, except for $\Lambda\to p$, where the muon-mode decay rate is taken from the recent LHCb measurement~\cite{LHCb:2025wld}, not yet included in the PDG.

The Standard Model prediction for the muon-to-electron mode ratio is then compared to the experimental value to determine constrains on non-standard scalar and tensor interactions, as depicted in Fig.~\ref{fig:cont}.
The orange contour shows the constrain coming from semileptonic hyperon decays using input from this work. We find
\begin{equation}
   \epsilon_S=-0.009(36)\qq{and}\epsilon_T= 0.011(21) \,,
\end{equation}
with a correlation of -0.61 in $\overline{\rm MS}$ at 2\,GeV.

The black contours indicate constraints derived from high-$p_T$
searches at the LHC, based on limits on non-standard $us\to e\nu$ interactions
in the process $pp\to e + \slashed{E}_T + X$ \cite{Cirigliano:2012ab, Chang:2014iba}. The individual curves correspond to different
center-of-mass energies $\sqrt{s}$ as indicated in the figure (with $\sqrt{s}$ = 7\,TeV for outer and 8\,TeV for inner line). It is worth mentioning that the Run~2 dataset collected at $\sqrt{s}=13$~TeV is expected to yield the strongest bound among the LHC contours, and accordingly its allowed region is smaller. Finally, the blue contour shows the constraint from semileptonic kaon decays
($K_{\ell3}$) from Ref.~\cite{Gonzalez-Alonso:2016etj}, which provides an additional
low-energy probe of the same effective interactions and currently is the most stringent constraint. In the $c\to s$ sector, it is also similarly constrained~\cite{Becirevic:2026tle}.

\section{Conclusions}\label{sec:concl}

We presented a lattice QCD determination of the scalar and tensor $\Lambda\to p$ transition form factors using an ensemble with physical pion mass. The calculation is performed using an ETMC $N_f=2+1+1$ twisted mass fermion  with lattice spacing $a\simeq 0.08$~fm and spatial volume (5.4 fm)$^3$.  Ground-state matrix elements are extracted using a two-state analysis to ensure a better determination of the ground-state matrix elements. The full $q^2$-dependence is described using a constrained $z$-expansion. Since the calculation is performed at a single  lattice spacing, discretization effects cannot be quantified within the present setup. A continuum extrapolation using additional lattice spacings is required to include these effects in the systematic uncertainty.

The resulting scalar and tensor form factors provide the missing non-perturbative input required for a controlled analysis of non-standard interactions in semileptonic hyperon decays. In particular, they enable a quantitative assessment of scalar and tensor contributions to the muon-to-electron decay-rate ratio
\(
R^{\mu e}=\Gamma(\Lambda\to p\mu\bar\nu_\mu)/\Gamma(\Lambda\to pe\bar\nu_e),
\)
an observable that is largely insensitive to CKM uncertainties and free of hadronic form-factor dependence in the Standard Model at next-to-leading order. Using our lattice QCD inputs together with the recent LHCb measurement of $R^{\mu e}$, we determine updated sensitivity coefficients and obtain revised constraints on the scalar and tensor  coefficients $(\epsilon_S,\epsilon_T)$ that probe physics beyond the Standard Model.

Compared to earlier analyses based on symmetry arguments and model estimates, the present work replaces phenomenological inputs for the $\Lambda\to p$ scalar and tensor form factors with lattice QCD results with quantified uncertainties and allowing for a non-perturbative treatment of decay rates. When combined with constraints from other hyperon channels, semileptonic kaon decays, and high-$p_T$ searches at the LHC, the allowed region in the $(\epsilon_S,\epsilon_T)$ plane is restricted to lie close to the Standard Model point.

\section*{ACKNOWLEDGMENTS}
We thank the members of the ETM Collaboration for their constructive cooperation. We thank Matteo Di Carlo, Antonio Evangelista, Roberto Frezzotti,
Giuseppe Gagliardi and Vittorio Lubicz, for useful discussions and
crosschecks on the analysis of the renormalization factors. S.B. and A.K. acknowledge support from \texttt{POST-DOC/0524/0001} and \texttt{VISION ERC - PATH 2/0524/0001}, and C.A. from \texttt{IMAGE-N Excellence hub/0524/0459}, co-financed by the European Regional Development Fund and the Republic of Cyprus through the Research and Innovation Foundation within the framework of the Cohesion Policy Programme “THALIA 2021-2027”. A.K. acknowledges financial support from ``The three-dimensional structure of the nucleon from lattice QCD'' \texttt{3D-N-LQCD} program, funded by the University of Cyprus.
Ensemble generation employed the open-source packages tmLQCD~\cite{Jansen:2009xp,Abdel-Rehim:2013wba,Deuzeman:2013xaa,Kostrzewa:2022hsv}, DD-$\alpha$AMG~\cite{Frommer:2013fsa,Alexandrou:2016izb,Bacchio:2017pcp,Alexandrou:2018wiv}, and QUDA~\cite{Clark:2009wm,Babich:2011np,Clark:2016rdz}. Analysis utilized the open-source software PLEGMA and QUDA. 
We gratefully acknowledge the Gauss Centre for Supercomputing e.V. (www.gauss-centre.eu) for computing time on JUWELS Booster at the Jülich Supercomputing Centre (JSC). We also acknowledge the Swiss National Supercomputing Centre (CSCS) and the EuroHPC Joint Undertaking for access to the Daint-Alps supercomputer. We are grateful to CINECA and the EuroHPC JU for access to the supercomputing facilities hosted at CINECA and Leonardo-Booster.

\begin{widetext}
\appendix
\section{Derivation of the sensitivity coefficients}
\label{app:sensitivity}

In this appendix we outline the derivation of Eq.~\eqref{Eq:sens} for the
scalar sensitivity coefficient. The tensor case follows an analogous
procedure. Our goal is to expand the decay rate at NLO in the small dimensionless parameter $\delta$, see Eq.~\eqref{eq:params}. The same expansion is commonly used for the electron-mode of the standard model decay rate~\cite{Cabibbo:2003cu}, yielding
\begin{equation}\label{eq:decay_LL_exp}
\Gamma^e_{LL}
=
\frac{G_F^2 |V_{us}|^2 m_\Lambda^5 \deltaLP^5}
{60\pi^3}
\Big[\left(1-\tfrac{3}{2}\deltaLP\right)
(f_1^2+3g_1^2)+\mathcal{O}\!\left(\deltaLP^2\right)\Big].
\end{equation}

For the scalar–vector interference contribution, we begin from the decay
rate defined as integral of its differential form in Eq.~\eqref{eq:decay_rate_LS},
\begin{align}
\Gamma_{LS} &=
\int_{m_\ell^2}^{q^2_{\rm max}}
\frac{G_F^2 |V_{us}|^2 \sqrt{s_+ s_-}}
{64\pi^3 m_\Lambda^3}
\left(1-\frac{m_\ell^2}{q^2}\right)^2
m_\ell\, s_+\, F_S(q^2)
\Bigg[
(m_\Lambda-m_p)F_1(q^2)
+
\frac{q^2}{m_\Lambda}F_3(q^2)
\Bigg]
dq^2 ,
\end{align}
where $s_\pm=(m_\Lambda\pm m_p)^2-q^2$. $F_3(q^2)$ is a second-class form factor, i.e. it is suppressed by the initial–final state mass splitting and is therefore $\mathcal{O}(\delta)$. Combined with its additional suppression by $q^2$ in the decay rate, it is negligible in this expansion. Moreover, the $q^2$ dependence of the remaining form factors can be neglected, entering at $\mathcal{O}(\delta^2)$, so that they may be approximated by their forward-limit values, $F_S(q^2)\approx f_S$ and $F_1(q^2)\approx f_1$. Finally, it is convenient to introduce the dimensionless variable
\begin{equation}
    x\equiv\frac{q^2}{(m_\Lambda+m_p)^2}\qquad\Longrightarrow\qquad s_+=(m_\Lambda+m_p)^2(1-x)\,,\quad s_-=(m_\Lambda-m_p)^2-x(m_\Lambda+m_p)^2
\end{equation}
and
\begin{equation}
\int_{m_\ell^2}^{q^2_{\rm max}}dq^2\rightarrow\int_{x_{\min}}^{x_{\max}}\!\!\!(m_\Lambda+m_p)^2dx \qq{with}
    x_{\rm min}\equiv\frac{m_\ell^2}{(m_\Lambda+m_p)^2}\approx0.3\% \qq{and} x_{\rm max}\equiv\frac{\Delta^2}{(m_\Lambda+m_p)^2}\approx0.7\%\,,
\end{equation}
thus being below percent level throughout the integral integral. Following this prescription, we obtain 
\begin{align}
\Gamma_{LS} &=
\frac{G_F^2 |V_{us}|^2\,\deltaLP^2}
{64\pi^3 m_\Lambda}
m_\ell\, (m_\Lambda+m_p)^5 f_S\,f_1\, \int_{x_{\rm min}}^{x_{\rm max}}
I(x)\,dx ,
\end{align}
where the dimensionless $I(x)$ collects all the $x$-dependent terms, being
\begin{equation}
I(x) \equiv \frac{s_+\sqrt{s_+ s_-}}{(m_\Lambda+m_p)^3(m_\Lambda-m_p)}\left(1-\frac{m_\ell^2}{q^2}\right)^2 =
\sqrt{(1-x)^{3}}\sqrt{1-x\frac{(m_\Lambda+m_p)^2}{(m_\Lambda-m_p)^2}}
\left(1-\frac{x_{\rm min}}{x}\right)^2.
\end{equation}

After expanding $\sqrt{(1-x)^{3}}$ as $1+\mathcal{O}(x)$, the integral over $x$ can be evaluated analytically, obtaining
\begin{equation}
\int_{x_{\rm min}}^{x_{\rm max}} I(x)dx
=
\frac{2m_\Lambda^3\deltaLP^3}
{15 m_\ell (m_\Lambda+m_p)^2}
\Big[\Pi(\deltamP)
+
\mathcal{O}(\deltaLP^2)\Big],
\end{equation}
where the phase-space function is
\begin{align}
\Pi(\deltamP) &=
\frac{5}{2}\deltamP
\Big[
(2+13\deltamP^2)\sqrt{1-\deltamP^2}
-3\deltamP^2(4+\deltamP^2)
\operatorname{artanh}
\!\left(\sqrt{1-\deltamP^2}\right)
\Big].
\end{align}
Substituting this result into the expression for $\Gamma_{LS}$ yields
\begin{equation}
\Gamma_{LS}
=
\frac{G_F^2 |V_{us}|^2 \deltaLP^5}
{480\pi^3}
m_\Lambda^2
(m_\Lambda+m_p)^3
f_1 f_S
\Big[\Pi(\deltamP)+
\mathcal{O}(\deltaLP^2)\Big]
=
\frac{G_F^2 |V_{us}|^2 \deltaLP^5}
{60\pi^3}
m_\Lambda^5
\Big[\left(1-\tfrac{3}{2}\deltaLP\right)
f_1 f_S\,
\Pi(\deltamP)
+
\mathcal{O}(\deltaLP^2)\Big]\,
\end{equation}
where in the right-hand side we have used $(m_\Lambda+m_p) = m_\Lambda(2-\delta)$ and expanded it. This term is the same originating the factor $(1-\tfrac32 \delta)$ also in Eq.~\eqref{eq:decay_LL_exp}, thus both cancelling when taking the ratio $\Gamma_{LS}/\Gamma_{LL}$. This factor was not accounted for in Eq.~(10) of Ref.~\cite{Chang:2014iba}, where 
$(m_\Lambda+m_p)$ was approximated with $2m_\Lambda$ introducing $\mathcal{O}(\delta)$ effects.

\section{z-expansion parameters}\label{app:pars}

Tables \ref{tab:zexp_S} and \ref{tab:zexp_H} provide the fitting parameters of the z-expansion discussed in Subsection \ref{z:exp}.
\begin{table}[h]
\begin{minipage}{0.35\textwidth}
\begin{tabular}{ccc}
\hline\hline
        $\alpha^{F_S}_0$ & $\alpha^{F_S}_1$ & $\alpha^{F_S}_2$ \\
        \hline
        0.8178(55) & -4.60(11)  & 11.5(10)\\
\hline\hline
\end{tabular}
\end{minipage}
\begin{minipage}{0.60\textwidth}
\centering
\begin{align*}
\begin{blockarray}{cccc}
~\text{corr.}~~ & \alpha^{F_S}_0 & \alpha^{F_S}_1 & \alpha^{F_S}_2 \\
\begin{block}{c(ccc)}
\alpha^{F_S}_0&1.00 & -0.61 & 0.15  \\
\alpha^{F_S}_1 &-0.61 & 1.00 & -0.86 \\
\alpha^{F_S}_2 &0.15 & -0.86 & 1.00  \\
\end{block}
\end{blockarray}
\end{align*}
\end{minipage}
   \caption{Coefficients of the $z$-expansion for the scalar form factor, along with their corresponding correlation. The $z$-expansion is of fourth order, with $a_3$ set as $a_3\equiv-(a_0+a_1+a_2)$ so that the form factor vanishes as $q^2 \to -\infty$. }
    \label{tab:zexp_S}
\end{table}

\begin{table}[h]
\begin{minipage}{0.35\textwidth}
\begin{tabular}{ccc}
\hline\hline
        $\alpha^{T_1}_0$ & $\alpha^{T_1}_1$ & $\alpha^{T_1}_2$ \\
        \hline
        0.7000(46) & -0.655(64)  & -2.36(47)\\
        $\alpha^{T_2}_0$ & $\alpha^{T_2}_1$ & $\alpha^{T_2}_2$ \\
        \hline
        -1.71(10) & 8.0(1.6)  & -13.9(7.0) \\
        $\alpha^{T_3}_0$ & $\alpha^{T_3}_1$ & $\alpha^{T_3}_2$ \\
        \hline
        1.361(67) & -4.9(1.0)  & 3.5(4.6) \\
        $\alpha^{T_4}_0$ & $\alpha^{T_4}_1$ & $\alpha^{T_4}_2$ \\
        \hline
        -0.804(75) & 4.0(1.1)  & -6.2(5.0) \\
\hline\hline
\end{tabular}
\end{minipage}
\begin{minipage}{0.60\textwidth}
\centering
\begin{align*}
\begin{blockarray}{ccccccccccccc}
~\text{corr.}~~ & \alpha^{T_1}_0 & \alpha^{T_1}_1 & \alpha^{T_1}_2 & \alpha^{T_2}_0 & \alpha^{T_2}_1 & \alpha^{T_2}_2 & \alpha^{T_3}_0 & \alpha^{T_3}_1 & \alpha^{T_3}_2& \alpha^{T_4}_0 & \alpha^{T_4}_1 & \alpha^{T_4}_2 \\
\begin{block}{c(cccccccccccc)}
\alpha^{T_1}_0&1.00 & -0.02 & -0.16 & -0.19 & 0.09 & -0.05 & 0.09 & -0.02 & -0.01 & -0.14 & 0.06 & -0.02 \\
\alpha^{T_1}_1 &-0.02 & 1.00 & -0.77 & -0.03 & -0.04 & 0.08 & 0.07 & 0.01 & -0.05 & -0.06 & -0.01 & 0.05 \\
\alpha^{T_1}_2 &-0.16 & -0.77 & 1.00 & -0.00 & 0.07 & -0.13 & -0.02 & -0.05 & 0.12 & 0.02 & 0.05 & -0.11 \\
\alpha^{T_2}_0&-0.19 & -0.03 & -0.00 & 1.00 & -0.89 & 0.78 & -0.85 & 0.76 & -0.67 & 0.90 & -0.80 & 0.70 \\
\alpha^{T_2}_1&0.09 & -0.04 & 0.07 & -0.89 & 1.00 & -0.97 & 0.74 & -0.84 & 0.82 & -0.79 & 0.89 & -0.87 \\
\alpha^{T_2}_2&-0.05 & 0.08 & -0.13 & 0.78 & -0.97 & 1.00 & -0.64 & 0.81 & -0.84 & 0.68 & -0.86 & 0.89 \\
\alpha^{T_3}_0&0.09 & 0.07 & -0.02 & -0.85 & 0.74 & -0.64 & 1.00 & -0.89 & 0.78 & -0.94 & 0.83 & -0.73 \\
\alpha^{T_3}_1&-0.02 & 0.01 & -0.05 & 0.76 & -0.84 & 0.81 & -0.89 & 1.00 & -0.97 & 0.83 & -0.93 & 0.90 \\
\alpha^{T_3}_2&-0.01 & -0.05 & 0.12 & -0.67 & 0.82 & -0.84 & 0.78 & -0.97 & 1.00 & -0.73 & 0.90 & -0.93 \\
\alpha^{T_4}_0&-0.14 & -0.06 & 0.02 & 0.90 & -0.79 & 0.68 & -0.94 & 0.83 & -0.73 & 1.00 & -0.89 & 0.78 \\
\alpha^{T_4}_1&0.06 & -0.01 & 0.05 & -0.80 & 0.89 & -0.86 & 0.83 & -0.93 & 0.90 & -0.89 & 1.00 & -0.97 \\
\alpha^{T_4}_2&-0.02 & 0.05 & -0.11 & 0.70 & -0.87 & 0.89 & -0.73 & 0.90 & -0.93 & 0.78 & -0.97 & 1.00 \\
\end{block}
\end{blockarray}
\end{align*}
\end{minipage}
\caption{Same as Table \ref{tab:zexp_S}, but for the tensor form factors}

\label{tab:zexp_H}
\end{table}

\newpage

\end{widetext}

\bibliography{main}{}

\end{document}